\documentclass[useAMS,usenatbib]{mn2e}
\usepackage{epsfig}
\usepackage{amsmath}
\usepackage{rotating}

\newcommand{\be}{\begin{equation}}
\newcommand{\beq}{\begin{equation}}
\newcommand{\ba}{\begin{eqnarray}}
\newcommand{\ee}{\end{equation}}
\newcommand{\eeq}{\end{equation}}
\newcommand{\ea}{\end{eqnarray}}
\newcommand{\msun}{{\rm M_{\odot}}}

\newcommand{\apj}{ApJ}

\newcommand{\jcrit}{{J_{21}^{\rm crit}}}

\def\lsim{~\rlap{$<$}{\lower 1.0ex\hbox{$\sim$}}}

\def\gsim{~\rlap{$>$}{\lower 1.0ex\hbox{$\sim$}}}

\title[Keeping Protogalactic Gas H$_2$ Free]{Supermassive Black Hole
  Formation by Direct Collapse: Keeping Protogalactic Gas H$_2$--Free in Dark Matter Halos with Virial Temperatures $T_{\rm vir}\gsim 10^4$ K}
\author[C. Shang, G. L. Bryan and Z. Haiman]{Cien Shang$^{1}$\thanks{E-mail: cien@phys.columbia.edu, gbryan@astro.columbia.edu, zoltan@astro.columbia.edu},
  Greg L. Bryan$^{2}$  and Z. Haiman$^{2}$\\
  $^{1}$Department of Physics, Columbia University, 538 West 120th Street, New York, NY 10027\\
  $^{2}$Department of Astronomy, Columbia University, 550 West 120th Street, New York, NY 10027}

\begin{document}

\date{\today}
\pagerange{\pageref{firstpage}--\pageref{lastpage}} \pubyear{2009}

\maketitle

\label{firstpage}

\begin{abstract}
  In the absence of ${\rm H_2}$ molecules, the primordial gas in early
  dark matter halos with virial temperatures just above $T_{\rm
  vir}\gsim 10^4$K cools by collisional excitation of atomic H.
  Although it cools efficiently, this gas remains relatively hot, at a
  temperature near $T\sim 8000$ K, and consequently might be able to
  avoid fragmentation and collapse directly into a supermassive black
  hole (SMBH).  In order for ${\rm H_2}$--formation and cooling to be
  strongly suppressed, the gas must be irradiated by a sufficiently
  intense ultraviolet (UV) flux.  We performed a suite of
  three--dimensional hydrodynamical adaptive mesh refinement (AMR)
  simulations of gas collapse in three different protogalactic halos
  with $T_{\rm vir}\gsim 10^4$K, irradiated by a UV flux with various
  intensities and spectra.  We determined the critical specific
  intensity, $\jcrit$, required to suppress ${\rm H_2}$ cooling in
  each of the three halos.  For a hard spectrum representative of
  metal--free stars, we find (in units of $10^{-21}~{\rm
  erg~s^{-1}~Hz^{-1}~sr^{-1}~cm^{-2}}$) $10^4<J_{21}^{\rm crit}<10^5$,
  while for a softer spectrum, which is characteristic of a normal
  stellar population, and for which ${\rm H^-}$--dissociation is
  important, we find $30<J_{21}^{\rm crit}<300$ . These values are a
  factor of 3--10 lower than previous estimates. We attribute the
  difference to the higher, more accurate ${\rm H_2}$ collisional
  dissociation rate we adopted. The reduction in $\jcrit$
  exponentially increases the number of rare halos exposed to
  super--critical radiation.  When ${\rm H_2}$ cooling is suppressed,
  gas collapse starts with a delay, but it ultimately proceeds more
  rapidly. The infall velocity is near the increased sound speed, and
  an object as massive as $M\sim 10^5~\msun$ may form at the center of
  these halos, compared to the $M\sim 10^2~\msun$ stars forming when
  ${\rm H_2}$--cooling is efficient.
\end{abstract}

\section{Introduction}
\label{sec:introduction}

The discovery of very bright quasars, with luminosities $\ge
10^{47}~{\rm erg~s^{-1}}$, at $z\simeq 6$ in the Sloan Digital Sky
Survey (SDSS) suggests that some SMBHs as massive as a few times
$10^9~\msun$ already existed when the universe was less than 1 Gyr old
(see, e.g., Fan 2006 for a review). In principle, these large black
hole (BH) masses, inferred from the apparent luminosities, could have
been overestimated due to strong gravitational lensing and/or
beaming. However, no obvious sign of either effect was found in the
images or spectra of these quasars (Willott et al. 2003; Richards et
al. 2004).

Explaining how SMBHs with masses in excess of $10^9~\msun$ could
assemble within 1 Gyr presents some challenges. Perhaps the most
natural proposal is that they grow, by a combination of
Eddington--limited accretion and mergers, from the stellar--mass seed
BHs provided by the remnants of the first generation of massive,
metal-free stars (e.g., Haiman \& Loeb 2001).  Indeed, the initial
seed BHs, with masses of the order of their progenitor stars, $\sim
100~\msun$ (Abel et al. 2000, 2002; Bromm et al. 2002), are expected
to be present very early (at redshifts prior to $z\gsim 10$). There
are, however, a number of potential difficulties with this
scenario. First, the early seeds must accrete near the Eddington rate
for a Hubble time, without any prolonged interruption (Haiman \& Loeb
2001), which requires that the BHs are continuously surrounded by
dense gas (e.g. Turner 1991; Alvarez et al 2009).  However, early seed BHs are expected to
undergo frequent mergers, and the gravitational waves emitted during
the BH mergers impart a strong recoil to the coalesced BHs
(e.g. Pretorius 2005; Campanelli et al. 2006; Baker et al. 2006). The
typical velocity of this gravitational kick is expected to exceed
$\sim 100~{\rm km/s}$, which is significantly higher than the escape
velocity ($\lsim 10~{\rm km/s}$) from typical dark matter halos at
$z\sim 10$. BHs are therefore easily ejected, or at least displaced
from the dense nuclei of their host halos at high redshift,
interrupting their accretion (Haiman 2004; Yoo \& Miralda-Escud\'{e}
2004; Shapiro 2005, Volonteri \& Rees 2006; Blecha \& Loeb 2008;
Tanaka \& Haiman 2009).  Even if such disruptive kicks are avoided
(because mergers at early times may be rare and/or occur primarily
between unequal--mass BHs; Volonteri \& Rees 2006; Tanaka \& Haiman
2009), there remain two fundamental difficulties.  First, when the
effects of photoheating and radiation pressure are included, the
maximum allowed steady accretion rate is significantly reduced, at
least in spherical symmetry (e.g. Milosavljevic et al. 2009a),
suggesting that accretion must be intermittent, with a time--averaged
rate well below the Eddington--limit (e.g. Ciotti \& Ostriker 2001,
2007; Milosavljevic et al. 2009b).  Second, models in which sufficient
numbers of early BHs are able to accrete and grow to $10^9~\msun$ by
$z\approx 6$ tend to overproduce the abundance of $\approx 10^6~\msun$
BHs by several orders of magnitude (compared to the abundance inferred
from local observations). This requires a preferential suppression of
BH growth in low--mass halos, plausibly due to negative feedback
effects acting on these halos (Bromley et al. 2004; Tanaka \& Haiman
2009).

An alternative way of assembling SMBHs is through more rapid
(super--Eddington) accretion or collapse. In this family of models,
primordial gas collapses directly into a BH as massive as
$10^4-10^6~\msun$ (Oh \& Haiman 2002 [hereafter OH02]; Bromm \& Loeb
2003 [hereafter BL03]; Koushiappas et al. 2004; Lodato \& Natarajan
2006; Spaans \& Silk 2006; Begelman et al. 2006; Volonteri et
al. 2008), possibly onto a pre--existing smaller seed BH (Volonteri \&
Rees 2005), or through the intermediate state of a very massive star
(BL03).  Such a head-start evades problems encountered by the models
where SMBHs grow at the Eddington limit from stellar mass seeds.  A
necessary condition for such direct collapse models is that the
collapsing gas avoid fragmentation; otherwise, normal Pop III stars
would be produced.\footnote{Another necessary condition is for the gas
to loose angular momentum efficiently.  Other than the angular
momentum transfer occurring above the resolution of our simulations,
this topic will not be addressed in the present paper. See,
e.g. Begelman et al. (2006) for a discussion and for references.}  At
the density of $\sim 10^4 {\rm cm^{-3}}$ (the critical density for
${\rm H_2}$; see below), the Jeans mass is $M_J\approx
10^6~\msun\,(T/10^4~{\rm K})^{3/2}$.  The gas temperature $T$ depends
crucially on whether ${\rm H_2}$ cooling is efficient: $T\approx
100~{\rm K}$, achievable if ${\rm H_2}$ cooling is efficient, would
imply $M_J\approx 10^3~\msun$, so that PopIII stars might form,
whereas $T\approx 10^4~{\rm K}$, expected in the absence of ${\rm
H_2}$, would yield $M_J\approx 10^6~\msun$, suggesting that direct
collapse into a $M\approx 10^6~\msun$ SMBH may be feasible (OH02).

Numerical simulations have indeed shown that fragmentation is
inefficient when ${\rm H_2}$ cooling is absent (BL03; Regan \&
Haehnelt 2009a, 2009b).  However, in most models, the absence of ${\rm
H_2}$ was {\rm assumed}, rather than derived.  The notable exceptions
are Spaans \& Silk (2006), whose model does not require any explicit
${\rm H_2}$ destruction\footnote{Spaans \& Silk propose that over a
relatively narrow range of densities and hydrogen column densities,
the Ly$\alpha$ photons emitted by atomic H cooling are trapped within
the collapsing gas -- this prevents the temperature from falling below
$\sim 8,000$K, and keeps the ${\rm H_2}$ molecules collisionally
dissociated.}, and BL03, who performed simulations with an ${\rm
H_2}$--photodissociating Lyman--Werner background.  The absence of
${\rm H_2}$ molecules from protogalactic halo gas can be justified by
a sufficiently intense UV radiation, either in the Lyman--Werner
bands, directly photo--dissociating ${\rm H_2}$ (near a photon energy
of $\sim 12$eV) or photo--dissociating the intermediary ${\rm H^-}$
(photon energies $\gsim 0.76$eV).  The relevant criterion is that the
photodissociation timescale is shorter than the ${\rm H_2}$--formation
timescale; since generically, $t_{\rm diss}\propto J$ and $t_{\rm
form}\propto \rho$, the condition $t_{\rm diss}=t_{\rm form}$ yields a
critical flux that increases linearly with density, $J^{\rm crit}
\propto \rho$.  In ``minihalos'', with virial temperatures $T_{\rm
vir} < 10^4$K, the gas cannot cool in the absence of ${\rm H_2}$, the
densities remain low ($\sim 1 {\rm cm^{-3}}$; e.g. Mesinger et
al. 2006) and ${\rm H_2}$ can be dissociated even by a relatively
feeble UV flux. The critical value has been found to be $J_{21}\sim
0.1$ (Haiman et al. 1997; Machacek et al. 2001, 2003; Mesinger et
al. 2006, 2008; Wise \& Abel 2007; O'Shea \& Norman 2008; here and in the rest of the
paper, $J_{21}$ denotes the specific intensity just below $13.6$eV, in
the usual units of $10^{-21} {\rm
erg~cm^{-2}~sr^{-1}~s^{-1}~Hz^{-1}}$).  This value is much smaller
than the expected level of the cosmic UV background in the
Lyman-Werner bands near reionization (BL03)
\begin{equation}
J_{\rm bg}
\approx \frac{1}{f_{\rm esc}} \frac{hc}{4 \pi} \frac{N_{\gamma} Y_{\rm H} \rho_{\rm b}}{m_{\rm p}}
\end{equation}
or
\begin{equation}
J_{21} \approx 40
\left(\frac{N_\gamma}{10}\right) 
\left(\frac{f_{\rm esc}}{0.1}\right)^{-1}
\left(\frac{1+z}{11}\right)^3,
\label{eq:Jbg} 
\end{equation} 
where $f_{\rm esc}$ is the escape fraction of ionizing radiation,
$N_{\gamma}$ is the average number of photons needed to ionize a
hydrogen atom, $Y_{\rm H}=0.76$ is the mass fraction of hydrogen,
$m_{\rm p}$ is the proton mass, and $\rho_{\rm b}$ is the background
baryon density with $\Omega_b h^2=0.023$.

The critical intensity $\jcrit$ in larger halos, with virial
temperatures $T_{\rm vir}\ge 10^4~{\rm K}$, however, is much higher
(Omukai 2001, hereafter OM01; OH02; BL03). This is primarily because
the gas in these halos can cool via excitations of atomic H and reach
much higher densities, and because the ${\rm H_2}$ molecules can then
become self--shielding (OH02).  In particular, for halos with $T_{\rm
vir}\sim 10^4$K, the value has been estimated in one--zone models to
be $\jcrit \approx 10^{3}-10^{5}$ (OM01). This range covers different
assumed spectral shapes; in particular, a thermal spectrum with
$T_*=10^4-10^5$K (OM01).  Using three--dimensional smooth particle
hydrodynamics (SPH) simulations, BL03 find $\jcrit \gsim 10^{5}$ (for
$T_*=10^5$K) and $\jcrit \lsim 10^{3}$ (for $T_*=10^4$K), in agreement
with the one--zone results.

The UV background that illuminates collapsing $T_{\rm vir}\approx
10^4$K halos was likely established by the massive pop-III stars that
had formed in previous generations of minihalos.  In this case, the
background spectrum is likely to be closer to the $T_*=10^5$K case,
implying $\jcrit gsim 10^{5}$. Furthermore, the early minihalos are
expected to be easily self--ionized, with most of their ionizing
radiation escaping into the intergalactic medium, i.e.  $f_{\rm
esc}\approx 1$ (Kitayama et al. 2004; Whalen, Abel \& Norman 2004).
Equation~\ref{eq:Jbg} shows that, unless $f_{\rm esc}$ is much
smaller, and/or $N_\gamma$ is large ($f_{\rm esc}/N_\gamma \lsim
10^{-3}$), the mean cosmic background is unlikely to reach the
required critical value.  The background will have inevitable spatial
fluctuations, and a very small fraction $f(J>J^{\rm crit})$ of $10^4$
K halos that have an unusually close and bright neighbor may still see
a sufficiently high flux. Dijkstra et al. (2008) used a model for the
three--dimensional spatial clustering of halos to estimate this
fraction, and found $f(J>10^3) \sim 10^{-6}$, {\it with an exponential
dependence of this result on $\jcrit$}.\footnote{Alternatively, the
critical value could be established by sources internal to the halo,
e.g. by a vigorous phase of starburst (Omukai \& Yoshii 2003) or by an
accreting stellar seed BH (Volonteri \& Rees 2005). However, having
gone through star--formation already, the halo gas is unlikely to
still be metal--free, and is then likely to fragment into low--mass
stars, rather than collapsing directly into a SMBH (Omukai et
al. 2008).}

In this paper, we derive detailed estimates for $\jcrit$ based on a
suite of 3D hydrodynamic simulations.  Our motivation is
two--fold. First a UV intensity that exceeds $\jcrit$ is crucial for
the feasibility of direct SMBH formation models. Second, the existing
estimates of $\jcrit$ significantly exceed the expected value of the
mean cosmic background.  If these estimates are correct, then
$J>\jcrit$ will be experienced only by those rare halos that probe the
bright tail of the spatially fluctuating background $J$.  In this
case, even a small change in the value of $\jcrit$ can cause a large
change in the expected number of halos that can form SMBHs by direct
collapse.

Our paper adds to the earlier work of OM01, which estimated $\jcrit$
based on a one-zone model with a fixed, prescribed collapse dynamics,
that could not address fragmentation, and to the results of BL03, who
use three--dimensional simulations, but report only approximate upper
and lower limits on $\jcrit$ for a single halo.  In addition, we study
the behavior of the collapsing gas in detail as a function of
$J_{21}$, and obtain a rough estimate for the fluctuations in $\jcrit$
by following the collapse of three different halos. We also obtain a
quantitative estimate of the final collapsed central massive object,
based on the infall time--scales observed in each case.

The rest of this paper is organized as follows. In
\S~\ref{sec:numerical}, we describe the simulation setup in detail. In
\S~\ref{sec:results}, we present the range of $\jcrit$ values derived
from the simulations, and explain the underlying physics using a
one-zone model. We also estimate the fluctuations in $\jcrit$.  In
\S~\ref{sec:fate}, we discuss the ultimate fate of the halos, for
different values of $J_{21}$. Finally, we summarize the results and
present the conclusion of this work in \S~\ref{sec:conclusion}.


\section{Numerical Methodology}
\label{sec:numerical}
We use the Eulerian adaptive mesh refinement (AMR) code Enzo, which
has been tested extensively and is publicly available (Bryan 1999;
Norman \& Bryan 1999; O'Shea et al. 2004). Enzo uses an N-body
particle-mesh solver to follow dark matter dynamics, and an Eulerian
AMR method by Berger \& Colella (1989) to solve the hydrodynamical
equations for an ideal gas. This combination allows for high dynamic
range in gravitational physics and hydrodynamics. Nested grids are
used whenever higher resolution is needed. At each new refinement
level, a parent grid is replaced by a few smaller child grids.

The chemical composition of the gas is followed by solving the
non--equilibrium evolution of nine species: ${\rm H}$, ${\rm H^{+}}$,
${\rm He}$, ${\rm He^{+}}$, ${\rm He^{++}}$, ${\rm H^{-}}$, ${\rm
H^{+}_{2}}$, ${\rm H_{2}}$, and ${\rm e^{-}}$ (Abel et al. 1997; Abel
et al. 2000). Our reaction network did not include ${\rm HD}$ or other
molecules involving deuterium. This should make very little difference
to our results, since ${\rm HD}$ cooling only becomes important for
temperatures below a few hundred Kelvin (McGreer \& Bryan 2008). The
$\rm H_2$ radiative cooling function of Galli \& Palla (1998) is
employed to follow the temperature of the gas.

A few modifications were made to the publicly available Enzo
code. First, we added direct ${\rm H^{-}}$ photodissociation into
chemistry solver,
\begin{eqnarray}
{\rm H^-     +h\nu    \rightarrow H  +e^-   }.
\end{eqnarray}
As explained below, this reaction is important in determining the
value of $\jcrit$.  Second, we included the self--shielding of ${\rm
H_2}$ in the LW bands when computing ${\rm H_2}$ photodissociation
rate.  Specifically, the intensity in the LW band is multiplied by a
self-shielding factor $f_{\rm sh}$ given by Draine \& Bertoldi (1996),
\begin{eqnarray}
\label{eqn:fsh}
f_{\rm sh}={\rm min}\left[1,\left(\frac{N_{\rm H_2}}{10^{14} {\rm cm}^{-2}}\right)^{-3/4}\right],
\end{eqnarray}
where $N_{\rm H_2}$ is the ${\rm H_2}$ column density. Since $N_{\rm
H_2}$ is a non--local quantity, it is computationally very expensive
to obtain its exact value. To save computing time, we made the
commonly used approximation,
\begin{eqnarray}
\label{eqn:nh2}
N_{\rm H_2}=f_{\rm H_2} n_{\rm tot} \lambda_{\rm J},
\end{eqnarray}
where $f_{\rm H_2}$, $n_{\rm tot}$ and $\lambda_{\rm J}$ are the ${\rm
H_2}$ fraction by number, the total particle number density, and the
Jeans length, respectively. With this approximation, $N_{\rm H_2}$ and
$f_{\rm sh}$ can be computed from the local values of the temperature,
density and ${\rm H_2}$ fraction. We will quantify the accuracy of
this approximation in \S~\ref{subsec:selfshielding} below.

The simulation is set up in a comoving box of size $1~h^{-1}~{\rm
Mpc}$, assuming a standard ${\rm \Lambda CDM}$ model with the
following parameter: $\Omega_{\rm DM}=0.233$, $\Omega_{b}=0.0462$,
$\Omega_{\Lambda}=0.721$, $\sigma_8=0.817$, $n_s=0.96$ and $h=0.701$
(Komatsu et al. 2009).  We first perform a preliminary run in order to
identify halos suitable for detailed study. This run has a root grid
with a resolution of $128^3$ and no nested grids. Radiative cooling is
turned off, so that the gas in the halos is unable to contract to high
densities in this run. We evolved the simulation to $z=10$, stopped
it, and used the HOP halo finding algorithm (Eisenstein \& Hut 1998)
to identify dark matter halos in the output files. Three halos, which
are labeled below as A, B and C, were chosen for high--resolution
re--runs. All three of these were selected to have virial masses of a
few $\times 10^7~\msun$ at a redshift of $z=10$. In this paper, virial
mass is defined to be the total mass, including both gas and dark
matter components, inside a spherical averaged overdensity of 200 with
respect to the critical density of the universe.

We generated a new set of initial conditions for the three chosen
halos. Three nested grids with twice finer resolution were added, so
that the effective resolution of the innermost grid was $1024^3$,
resulting in a dark matter particle mass of 86 $\msun$.  Radiative
cooling was turned on in the re--runs, and the grid cells were
adaptively refined based on the following three criteria: baryon mass,
dark matter mass and Jeans length. According to the first two
criteria, additional grids are added when the baryon (dark matter)
mass in a grid cell exceeds 68 (683) solar mass, corresponding to 4
(8) times the initial mass in one grid cell (particle) in the most
refined region.  The third criterion ensures that the Jeans length is
resolved by at least 4 grid cells, so that no artificial fragmentation
would take place.  In addition, to avoid numerical effects due to the
finite mass of dark matter particles, the gravity of dark matter
particles is smoothed at refinement level 13, which corresponds to a
smoothing scale of $0.954/h$ (comoving) parsec. Each dark matter
particle has a mass of $\sim 85~\msun$. We allow the simulations to
proceed until a maximum refinement level of 18 is achieved,
corresponding to a resolution of $0.0298/h$ (comoving) parsec, or
about 800 AU (absolute).

For each of the three halos, we ran a series of simulations with
different UV spectra and intensities. For the spectral shape, we
adopted a Planck spectrum with a black--body temperature of either
$T_*={\rm 10^4~K}$ or $T_*={\rm 10^5~K}$ (hereafter denoted by T4 and
T5, respectively).  The softer of these spectra is meant to
approximate the mean spectrum of a normal stellar population, whereas
the higher--temperature case is closer to the harder spectrum expected
to be emitted by the first generation of massive, metal--free stars
(Tumlinson \& Shull 2000; Bromm, Kudritzki \& Loeb 2001; Schaerer
2002).  Using these two spectral types allows us to compare our
results with previous work (OM01, BL03) which adopted the same
spectral shapes.  In Table \ref{tbl:t4} (\ref{tbl:t5}), we list the
redshift ($z_{\rm col}$), virial mass ($m_{\rm vir,col}$) and central
gas temperature ($T_{\rm cent}$) in the halos when their cores
collapse in the presence of type T4 (T5) UV background.  Here, ``core
collapse'' is simply defined as the time when the maximum refinement
level (level 18) is reached.  In practice, once the collapse starts,
it proceeds very rapidly. As a result, the refinement level adopted
for this definition makes little difference to our results, as long as
it is chosen to be at level 13 or higher.  We generally varied
$J_{21}$ by factors of 10, but included additional runs with $J_{21} =
3 \times 10^2$ for halos B and C in the T4 case in order to determine
$\jcrit$ more precisely.  The values for the $J_{21}=10^3$ case for
halo C are missing from Table \ref{tbl:t4}, because the halo moved out
of the refinement region before it collapsed in this run.  This
occurred because of the late collapse redshift for this halo (recall
that halos were selected at $z=10$).  We could have rerun the
simulation with a larger refined region, but this was unnecessary
because the $J_{21} = 3 \times 10^2$ run was sufficient for
determining $\jcrit$.
 
\begin{table*}
  \caption{Redshift ($z_{\rm col}$), virial mass ($m_{\rm vir,col}$)
and central gas temperature ($T_{\rm cent}$) of the halos when their
cores collapse in the presence of a $T_*=10^4$K black--body (``T4'')
UV background.}
    \label{tbl:t4}
\begin{center}
\begin{tabular}{l c c c c c c c c c}
\hline \hline
&\multicolumn{3}{c}{Halo A}&\multicolumn{3}{c}{Halo
  B}&\multicolumn{3}{c}{Halo C}\\
$J_{21}$&$z_{\rm col}$&$m_{\rm vir,col}$($\msun$)&$T_{\rm cent}$(K)&$z_{\rm
  col}$&$m_{\rm vir,col}$($\msun$)&$T_{\rm cent}$(K)&$z_{\rm col}$&$m_{\rm
  vir,col}$($\msun$)&$T_{\rm cent}$(K) \\
\hline
$ 1 \times 10^{0}$ & 15.47 & $ 6.87 \times 10^{6}$ & 695.18 & 9.63 & $ 6.76 \times 10^{7}$ & 879.89 & 8.48 & $ 7.55 \times 10^{7}$ & 888.19 \\
$ 1 \times 10^{1}$ & 12.22 & $ 2.49 \times 10^{7}$ & 900.80 & 9.53 & $ 6.94 \times 10^{7}$ & 778.41 & 7.89 & $ 8.85 \times 10^{7}$ & 951.86 \\
$ 3 \times 10^{1}$ & 11.41 & $ 3.16 \times 10^{7}$ & 955.52 & 9.11 & $ 8.80 \times 10^{7}$ & 957.48 & 7.81 & $ 9.08 \times 10^{7}$ & 974.82 \\
$ 1 \times 10^{2}$ & 10.02 & $ 5.33 \times 10^{7}$ & 5985.60 & 8.93 & $ 9.59 \times 10^{7}$ & 807.96 & 7.65 & $ 9.77 \times 10^{7}$ & 890.45 \\
$ 3 \times 10^{2}$ & - & - & - & 8.91 & $ 9.63 \times 10^{7}$ & 6356.30 & 7.61 & $ 1.04 \times 10^{8}$ & 6334.20 \\
$ 1 \times 10^{3}$ & 9.96 & $ 5.43 \times 10^{7}$ & 6291.10 & 8.90 & $
9.67 \times 10^{7}$ & 6395.90 & - & - & - \\
\hline
\end{tabular}
\end{center}
\end{table*}

\begin{table*}
  \caption{Same as Table~\ref{tbl:t4}, but for a $T_*=10^5$K black--body (``T5'') UV background.}
    \label{tbl:t5}
\begin{center}
\begin{tabular}{l c c c c c c c c c}
\hline \hline
&\multicolumn{3}{c}{Halo A}&\multicolumn{3}{c}{Halo
  B}&\multicolumn{3}{c}{Halo C}\\
$J_{21}$&$z_{\rm col}$&$m_{\rm vir,col}$($\msun$)&$T_{\rm cent}$(K)&$z_{\rm
  col}$&$m_{\rm vir,col}$($\msun$)&$T_{\rm cent}$(K)&$z_{\rm col}$&$m_{\rm
  vir,col}$($\msun$)&$T_{\rm cent}$(K) \\
\hline
$ 1 \times 10^{0}$ & 19.75 & $ 1.29 \times 10^{6}$ & 955.27 & 19.57 & $ 5.70 \times 10^{5}$ & 769.93 & 11.23 & $ 7.97 \times 10^{6}$ & 718.78 \\
$ 1 \times 10^{2}$ & 12.27 & $ 2.45 \times 10^{7}$ & 973.94 & 9.65 & $ 6.74 \times 10^{7}$ & 845.56 & 8.30 & $ 7.89 \times 10^{7}$ & 918.84 \\
$ 1 \times 10^{4}$ & 10.03 & $ 5.31 \times 10^{7}$ & 1055.20 & 8.94 & $ 9.56 \times 10^{7}$ & 963.06 & 7.49 & $ 1.08 \times 10^{8}$ & 998.76 \\
$ 3 \times 10^{4}$ & 9.99 & $ 5.39 \times 10^{7}$ & 6329.20 & 8.92 & $ 9.65 \times 10^{7}$ & 1064.40 & 7.07 & $ 1.23 \times 10^{8}$ & 988.36 \\
$ 1 \times 10^{5}$ & 9.93 & $ 5.49 \times 10^{7}$ & 6368.90 & 8.90 & $ 9.68 \times 10^{7}$ & 6446.30 & 7.48 & $ 1.09 \times 10^{8}$ & 6322.90 \\
\hline
\end{tabular}
\end{center}
\end{table*}

\section{The critical value of $J_{21}$}
\label{sec:results}

\subsection{Results from Three--Dimensional Simulations}
\label{subsec:enzoresults}

\begin{figure}
\begin{tabular}{c}
\rotatebox{-0}{\resizebox{90mm}{!}{\includegraphics{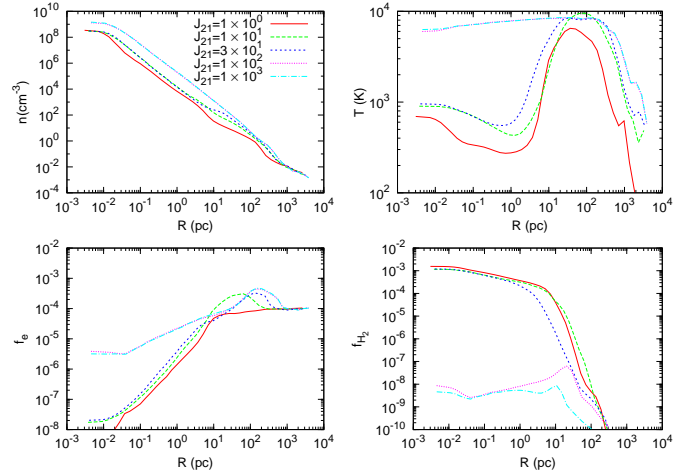}}}
\end{tabular}
\caption{Spherically averaged profiles of the particle density, gas
  temperature, ${\rm e^{-}}$ fraction and ${\rm H_2}$ fraction in halo
  A, for different values of the intensity $J_{21}$ of a type T4
  background.}
\label{fig:at4}
\end{figure}

\begin{figure}
\begin{tabular}{c}
\rotatebox{-0}{\resizebox{90mm}{!}{\includegraphics{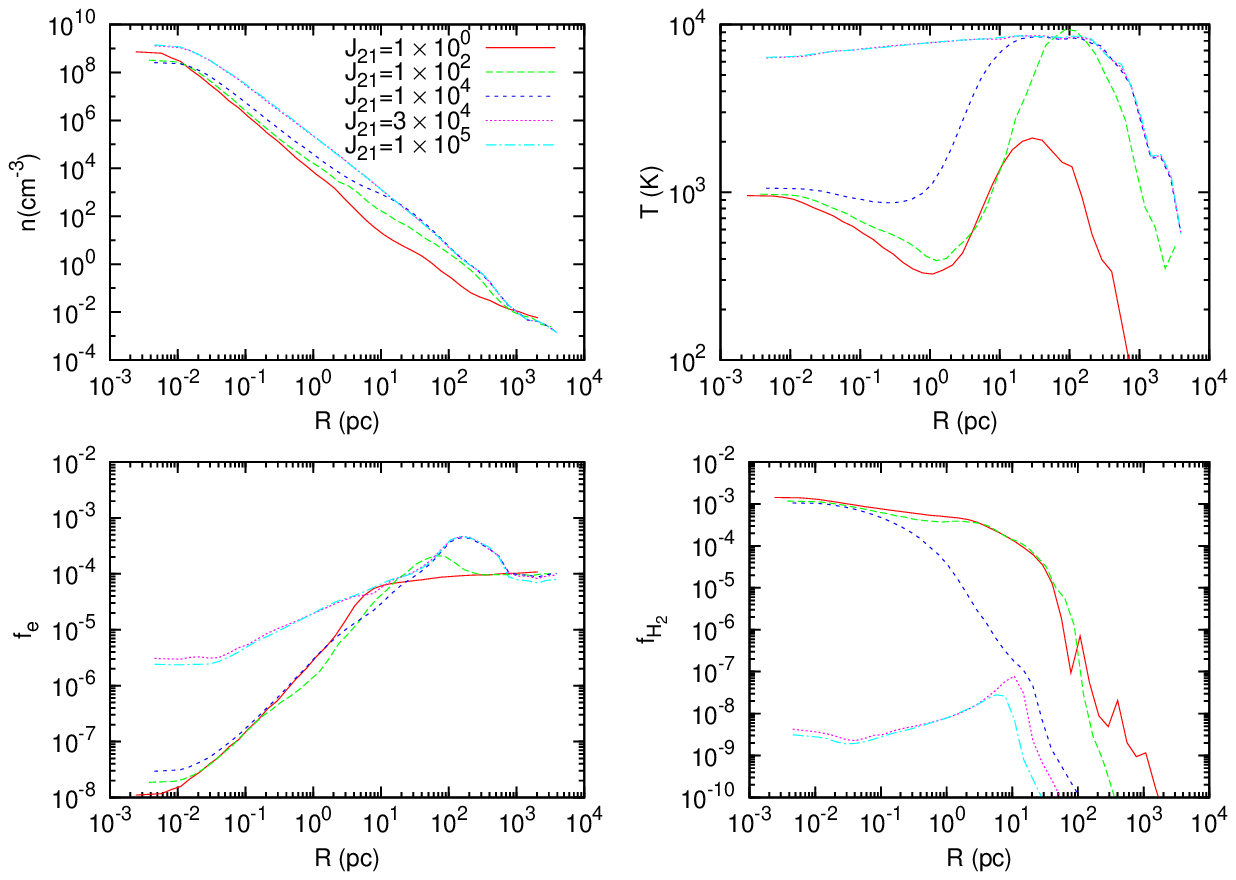}}}
\end{tabular}
\caption{Same as Figure \ref{fig:at4} except for type T5 backgrounds.}
\label{fig:at5}
\end{figure}

\begin{figure}
\begin{tabular}{c}
\rotatebox{-0}{\resizebox{90mm}{!}{\includegraphics{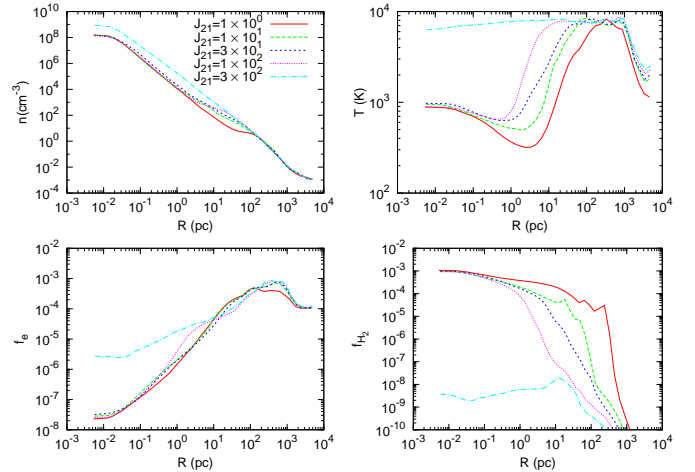}}}
\end{tabular}
\caption{Same as Figure \ref{fig:at4} except for Halo C.} 
\label{fig:bt4}
\end{figure}

\begin{figure}
\begin{tabular}{c}
\rotatebox{-0}{\resizebox{90mm}{!}{\includegraphics{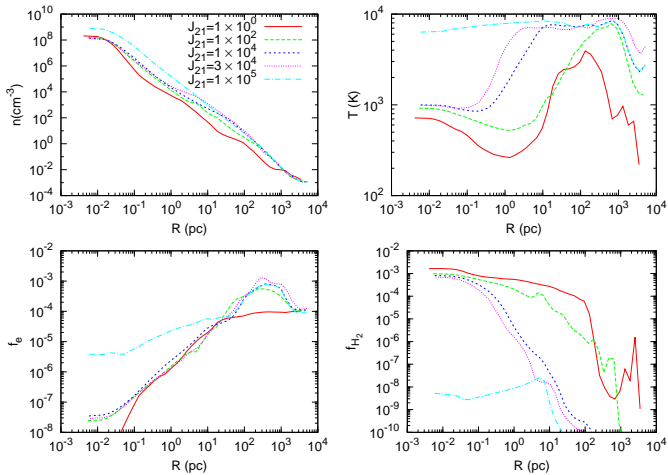}}}
\end{tabular}
\caption{Same as Figure \ref{fig:bt4} except for type T5 backgrounds.} 
\label{fig:bt5}
\end{figure}

Based on the central gas temperatures ($T_{\rm cent}$) listed in
Tables \ref{tbl:t4} and \ref{tbl:t5}, the halos can be grouped
unambiguously into two categories: ``cool'' halos and ``hot''
halos. The ``cool'' halos all have central temperatures near $\sim
1000$ K, whereas the ``hot'' halos all have central temperatures near
$\sim 6000$ K. To see the differences between these two categories
more clearly, we show in Figures \ref{fig:at4} - \ref{fig:bt5} the
spherically averaged profiles of the particle density, gas
temperature, ${\rm e^{-}}$ fraction and ${\rm H_2}$ fraction in halos
A and C, for different values of $J_{21}$, and in the presence of type
T4 and T5 UV backgrounds. The profiles of halo B are not shown because
they are very similar to those of halo C. All profiles are shown at
the time of collapse at $z_{\rm coll}$. The collapse is delayed
significantly as the intensity of the background is increased (see
Tables \ref{tbl:t4} and \ref{tbl:t5}), so that the curves shown in
each panel reflect conditions at significantly different redshifts.

Consistent with the values in Tables \ref{tbl:t4} and \ref{tbl:t5},
these figures show that central temperatures are very different for
``hot'' and ``cool'' halos. However, the temperature profiles at large
radii are actually similar in the cases where $J_{21}\ge 10$.  As the
gas falls in, the temperature first rises due to virialization shocks,
and stays at $\sim 8000$ K because of atomic cooling. The clear
differences arise around $R \sim 10~{\rm pc}$, where the ``cool''
halos start to cool via ${\rm H_2}$, while the ``hot'' halos stay hot.
As the figures show, $R \sim 10~{\rm pc}$ corresponds to a density of
$n\sim 10^3~{\rm cm^{-3}}$. The conditions at this density are crucial
in determining the value of $\jcrit$.  In particular, the difference
in the temperature profiles is clearly induced by the difference in
the ${\rm H_2}$ fractions. As the bottom right panel of each of
Figures \ref{fig:at4} - \ref{fig:bt5} shows, the ${\rm H_2}$ fractions
increase with decreasing radii in the ``cool'' halos. In the ``hot''
halos, the fractions increase with decreasing radii outside $R\sim
10~{\rm pc}$, but the fraction stays constant, at the low level of
$\sim 10^{-8}$, inside this radius.  There are also clear differences
in ${\rm e^{-}}$ fraction profiles: the cores of the ``hot'' halos
have a much larger electron fraction ($\sim 10^{-5}$), whereas in the
cores of the ``cool'' halos, the electrons are depleted.  Finally,
although the density profiles are overall quite similar, the gas
densities reached within the inner $\sim 10~{\rm pc}$ in the ``hot''
halos are noticeably higher than in the ``cool'' halos.

As we will discuss in more detail in \S~\ref{sec:fate} below, during
the process of quasi--static contraction, the density and temperature
adjust so that the cooling, sound crossing, and dynamical time scales
are all approximately equal ($t_{\rm cool} \approx t_{\rm cross}
\approx t_{\rm ff}$).  From the condition $t_{\rm cross} \approx
t_{\rm ff}$, we find $nR^2 \propto T$ (see equation~\ref{eqn:vtheorem}
below), which explains why the density is higher when temperature is
higher at a given radius.
Within $R\sim 10~{\rm pc}$, the ${\rm e^{-}}$ fractions in ``hot''
halos have a much shallower slope than in ``cool'' halos.  This is
because the main reactions determining the electron fraction
(reactions 1 and 4 in the Appendix) are highly temperature sensitive
-- in particular, recombinations (reaction 4) are exponentially more
rapid at lower temperatures, and the gas is therefore much more fully
recombined in the ``cool'' halos.

It is worth noting that in the case of $J_{21}=1$ and type T5 UV
background, the halo gas is able to collapse before the halo potential
grows deep enough to reach a virial temperature of $\sim 8000$ K,
where atomic cooling becomes important.  Effectively, at this low UV
background, the halos are ``minihalos'', whose properties are fully
determined by ${\rm H_2}$ cooling.  This causes a ``break'' in the
properties of halos, such as their mass and collapse redshift, between
$J_{21}=1$ and $J_{21}=10$. Such a ``break'' was also seen in O'Shea
\& Norman (2008), but at $J_{21}\sim 0.1$. The fact that this break
occurs at a higher $J_{21}$ in our simulations is not surprising, since
we include self-shielding, which was ignored in O'Shea \& Norman
(2008).

Finally, we note that even in the cases with higher values of
$J_{21}$, when the collapse is delayed, and the dark matter halo
potentials grow somewhat deeper, the effective virial temperature of
the halos we examined are still only marginally (if at all) above
$10^4$K.  Indeed, Figures \ref{fig:at4} - \ref{fig:bt5} show that the
gas in the simulated halos is never fully collisionally ionized -- the
highest free electron fraction reached is only $10^{-3}$.  This regime
differs from the common assumption that ``second generation'' halos
are collisionally ionized (e.g. OH02).  In the strict absence of any
${\rm H_2}$ cooling, the shocks that occur in more massive halos
could, of course, eventually produce full ionization.  However, prior
to building up this higher mass, every halo must go through the
``marginal'' stage where its effective virial temperature is close to,
but just above, $T_{\rm vir}\approx 10^4$K.  As our results show, the
gas at this stage can already cool efficiently via excitations of
atomic H.  Therefore, we expect that halos that reach virial
temperatures significantly above $\approx 10^4$K and become fully
ionized, but still consist of pure atomic H (and He), will be
exceedingly rare.

Because each of our runs produce a collapsed halo that belongs
unambiguously either to the ``cool'' or ``hot'' category, we simply
define $\jcrit$ as the value that divides these two regimes.  Our
suite of simulations is then sufficient to determine $\jcrit$ to
within a factor $\sim 3$ for each of the three halos, and for the two
types of UV spectra. The results are listed in Table \ref{tbl:jcrit}.
\begin{table}
  \caption{$\jcrit$ for the three different halos (A, B,
  and C), and for the two different UV background spectra (T4 and
  T5).}
    \label{tbl:jcrit}
\begin{center}
\begin{tabular}{l c c c c}
\hline \hline
 & Halo A & Halo B & Halo C & One-zone\\
\hline
T4& $3\times 10^{1}-10^2$&$10^2-3\times 10^{2}$&$10^2-3\times
10^{2}$ & $3.9\times10^1$ \\
T5& $ 10^{4}-3\times 10^{4}$&$3\times 10^{4}-10^{5}$&$3\times
10^{4}-10^{5}$ & $1.2\times 10^4$\\ 
\hline
\end{tabular}
\end{center}
\end{table}

There are two interesting new points that can be concluded from the
$\jcrit$ values in Table \ref{tbl:jcrit}.  First, the ranges of
$\jcrit$ we found are smaller than previously estimated. Previous
studies (OM01; Omukai et al. 2008) found $\jcrit\approx 10^{3}$ for
the type T4 spectrum and $\sim 3\times 10^{5}$ for the type T5
spectrum. The upper and lower limits of $\lsim 10^{3}$ and $\gsim
10^{5}$, reported from SPH simulations by BL03 for these two types of
spectra, are consistent with the above values.  These values, however,
are a factor of $\sim 10$ larger than our results for halo A, and a
factor of $\sim 3$ larger than for halos B and C. Second, even with
our crude sampling of $J$ values, we see a ``scatter'' in $\jcrit$:
the critical flux is a factor of $\sim 3$ lower for halo A than for
halos B and C, despite the similar masses and collapse redshifts of
these halos.

The fact that we find $\jcrit$ values that are smaller than previously
estimated is particularly important, because the critical UV fluxes
are high compared to the expected level of the cosmic UV background at
high redshifts. As noted above, however, the background will
inevitably have spatial fluctuations, and a small fraction of halos
may still see a sufficiently high flux. Dijkstra et al. (2008) used a
model for the three--dimensional spatial clustering of halos to
estimate the probability distribution function (PDF) of the background
LW flux $J_{\rm bg}$, as sampled by DM halos with $T_{\rm vir}\approx
10^4$K. Their results show that the interesting range of $\jcrit$
samples the bright, steeply falling tail of the flux PDF.  In
particular, in their Figure 2, Dijkstra et al. (2008), show the
fraction $f(>\jcrit)$ of halos exposed to a flux above a given
$\jcrit$.  This fraction is very sensitive to $\jcrit$: for $\jcrit
\gsim 10^{4}$, the PDF drops to negligibly low values ($f\lsim
10^{-8}$); for $\jcrit=10^{3}$, $f\sim 10^{-6}$, whereas for
$\jcrit=10^{2}$, $f\sim 10^{-3}$.  This implies that the 10--fold
decrease we find in the value of $\jcrit$ increases the number of
candidate DM halos, where direct SMBH formation may be feasible, by a
factor of $\approx 10^3$.

\subsection{Results from One--Zone Models}
\label{subsec:onezone}

In order to understand the above results, we performed one-zone
calculations similar to those in OM01.  The chemical reaction and
cooling rates are set to be the same as in the Enzo runs, but the gas
density is assumed to have a single value that follows a fixed
prescribed evolution.  The list of chemical reactions we used, and
their rates, are shown in the Appendix. In this model, the dark matter
and the baryons start their collapse at the turnaround redshift, which
is set to be $z=17$.  At this redshift, zero initial velocities are
assumed.  The subsequent evolution of the dark matter density is
computed using the spherical collapse model, up to the time of
virialization, after which it is assumed to stay constant at the
virial density. The evolution of the baryonic component is followed
using the equation,
\begin{eqnarray}
\label{eqn:gasevl}
\frac{{\rm d}\rho_{b}}{{\rm d}t}=\frac{{\rm d}\rho_{b}}{t_{\rm ff}},
\end{eqnarray}
where $\rho_b$ is the baryonic density, and $t_{\rm
ff}=\sqrt{3\pi/32G\rho}$ is the dynamical (free-fall) time.  The
thermal evolution is described the following equation,
\begin{eqnarray}
\label{eqn:themalevl}
\frac{{\rm d}e}{{\rm d}t}=-p\frac{{\rm d}}{{\rm
    d}t}\left(\frac{1}{\rho_b}\right)-\frac{\Lambda_{\rm net}}{\rho_b},
\end{eqnarray}
where $e$ is the internal energy per unit baryonic mass, $p$ is the
gas pressure, and $\Lambda_{\rm net}$ is the net cooling rate computed
using cooling function by Galli and Palla (1998). The internal energy
density is computed from
\begin{eqnarray}
\label{eqn:e}
e=\frac{1}{\gamma -1}\frac{k_{\rm b}T}{\mu m_{\rm H}},
\end{eqnarray}
where $\gamma=5/3$ is the adiabatic index, $k_{\rm B}$ is the
Boltzmann constant, $\mu$ is the mean molecular weight, and $m_{\rm
H}$ is the mass of a hydrogen nucleus. We include ${\rm H_2}$
self-shielding in the same way as described above for the Enzo runs.

\begin{figure}
\begin{tabular}{c}
\rotatebox{-0}{\resizebox{90mm}{!}{\includegraphics{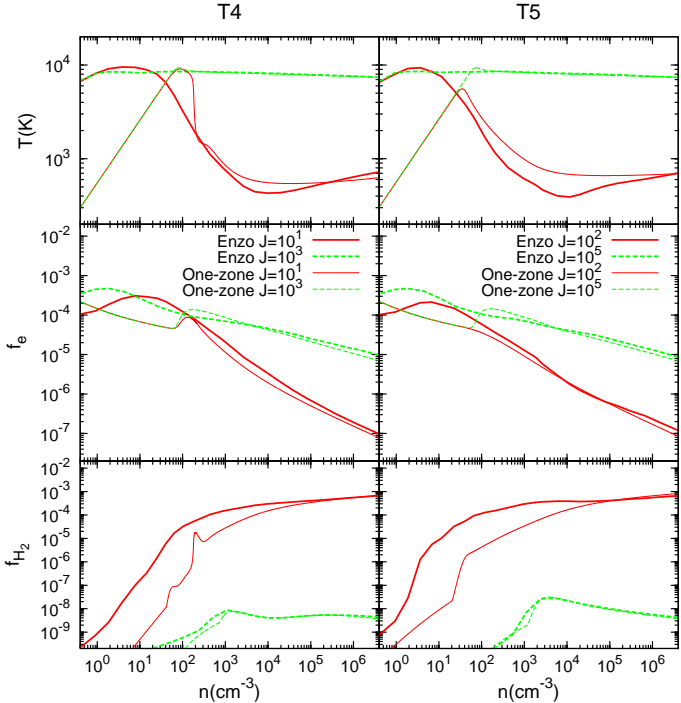}}}
\end{tabular}
\caption{Gas temperature, electron fraction and ${\rm H_2}$ fraction
 as a function of density computed for halo A with Enzo (thick curves)
 and in a one-zone model (thin curves). The solid and dashed curves
 show result for ``cool'' and ``hot'' halos, respectively, with the
 values of $J_{21}$ as indicated in the middle panels. The left panels
 are for type T4 UV backgrounds, and the right panels are for type T5
 UV backgrounds.}
\label{fig:compare}
\end{figure}

Figure \ref{fig:compare} shows the gas temperature, electron fraction
and ${\rm H_2}$ fraction as a function of density computed for halo A
with Enzo (thick curves) and in the one-zone model (thin curves) for
both type of UV backgrounds and for both ``cool'' halos (solid curves)
and ``hot'' halos (dashed curves), with $J_{21}$ values as indicated
in the middle panels.

The results from the simulation and the one-zone model are in
excellent agreement at $n\ge 10^{2}~{\rm cm^{-3}}$. There is a clear
difference in the temperatures at $n\le 10^{2}~{\rm cm^{-3}}$, which
is expected, since the one-zone model prescribes a smooth adiabatic
collapse, and the heating of the gas is purely due to this adiabatic
compression. In contrast, in the simulation, the gas experiences
shocks, which elevate the temperatures in the low--density regime.  To
check whether the shocks affect the comparison at higher densities, we
mimicked the shocks in the one--zone model by artificially setting the
temperature to $\sim $ 8000~{\rm K} at a fixed low density.  We found
that this makes the evolution match the simulations better at $n\le
10^{2}~{\rm cm^{-3}}$, but has little effect at higher density. In
particular, $\jcrit$ computed in the one--zone model (see below)
changes by $\le 3\%$.  This is because, as already mentioned above
(and will be discussed further below), $\jcrit$ is primarily
determined by the conditions at the critical density of ${\rm H_2}$,
which is $\sim 10^{3}~{\rm cm^{-3}}$.

We performed one--zone calculations similar to those shown in
Figure~\ref{fig:compare}, but varied $J$ according to a
Newton--Raphson scheme, until we converged on the critical value.  We
found $\jcrit=39$ for type T4 backgrounds, and $J_{21}^{\rm
crit}=1.2\times 10^{4}$ for type T5 backgrounds. These values fall
within the range of $\jcrit$ identified in the Enzo runs for halo A,
but are slightly below the Enzo range for halos B and C. We will
discuss possibly sources of this difference below.

\subsection{The importance of collisional dissociation}
\label{subsec:colldiss}

\begin{figure}
\begin{tabular}{c}
\rotatebox{-0}{\resizebox{90mm}{!}{\includegraphics{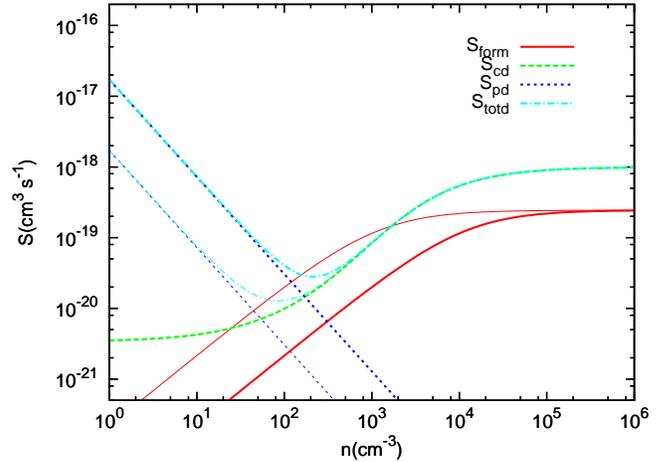}}}
\end{tabular}
\caption{The normalized reaction speed $S$ as a function of $n$ for
   type T4 backgrounds. {\it Solid curves}: formation speed $S_{\rm
   form}$. {\it Dotted curves}: photo--dissociation speed $S_{\rm
   pd}$. {\it Dashed curves}: collisional dissociation speed $S_{\rm
   cd}$. {\it Dash-dotted curves}: total dissociation speed $S_{\rm
   totd}$. The thin and thick curves are for $J_{21}=10^1$ and
   $J_{21}=10^2$, respectively. }
\label{fig:explainT4}
\end{figure}

\begin{figure}
\begin{tabular}{c}
\rotatebox{-0}{\resizebox{90mm}{!}{\includegraphics{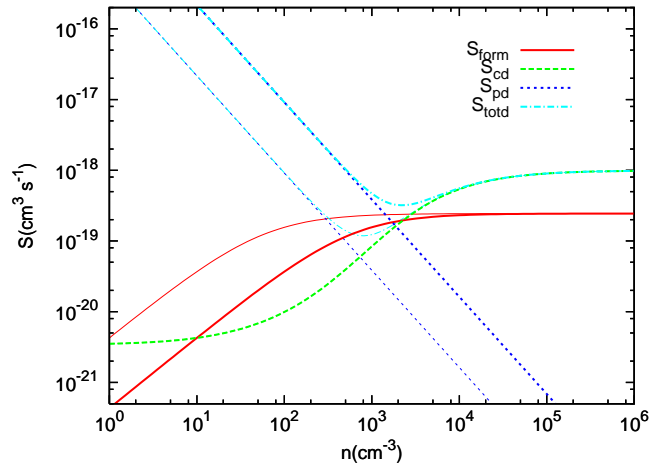}}}
\end{tabular}
\caption{$S$ as a function of $n$, as in Figure \ref{fig:explainT4},
  but for type T5 backgrounds. Line types are the same as in Figure
  \ref{fig:explainT4}, except here the thin and thick curves are for
  $J_{21}=10^4$ and $J_{21}=10^5$, respectively.}
\label{fig:explainT5}
\end{figure}

The one-zone model described above enables us to study the effect of
each reaction on $\jcrit$, thereby understanding the physics that
determines $\jcrit$, and identifying the uncertainty in $\jcrit$ due
to uncertainties in the chemical reaction rates.

The five most important reactions in determining $\jcrit$ are (OM01):
${\rm H_2}$ formation (reaction 10), ${\rm H^-}$ formation (reaction
9), ${\rm H^-}$ photo-dissociation (reaction 25), ${\rm H_{2}}$
photo-dissociation (reaction 28) and ${\rm H_{2}}$ collisional
dissociation (reaction 15). Reactions 9 \& 10 are the main channel of
forming ${\rm H_{2}}$ molecules. Reaction 25 competes with reaction 10
for ${\rm H^-}$, modulating the effective ${\rm H_{2}}$ formation rate
(OM01),
\begin{eqnarray}
\label{eqn:kform}
k_{\rm form}=k_9\frac{k_{10}n}{k_{10}n+k_{25}},
\end{eqnarray}
where $k_{\rm i}$ is the rate of reaction $i$ (as listed in the
Appendix). Both photo-dissociation and collisional dissociation
compete with this effective ${\rm H_2}$ formation process. The ${\rm
H_2}$ fraction in the gas is given approximately by the equilibrium
value for whichever dissociation process dominates,
\begin{eqnarray}
\label{eqn:fh2}
f_{\rm H_2}={\rm min}\left(\frac{k_{\rm form}}{k_{28}}f_{e}n,\frac{k_{\rm form}}{k_{15}}f_{e}\right).
\end{eqnarray}
To illustrate how these reactions determine the critical value of
$J_{21}$, in Figures \ref{fig:explainT4} and \ref{fig:explainT5} we
show the normalized reaction speed $S$ as a function of $n$ for
different values of $J_{21}$. The normalized formation,
photo-dissociation, collisional dissociation and total dissociation
speed are defined as
\begin{eqnarray}
\label{eqn:speed}
S_{\rm form}&=&k_{\rm form}n_{\rm e}n_{\rm H}/n^2\approx k_{\rm form}f_{\rm
  e},\\
S_{\rm pd}&=&k_{28}f_{\rm sh}n_{\rm H_2}/n^2=k_{28}f_{\rm sh}f_{\rm H_2}/n,\\
S_{\rm cd}&=&k_{15}n_{\rm H_2}n_{\rm H}/n^2\approx k_{15}f_{\rm H_2},\\
S_{\rm totd}&\approx&S_{\rm pd}+S_{\rm cd}.
\end{eqnarray}
In these figures, the temperature, $f_{\rm H_2}$ and $f_{\rm e}$ are
set to $8000~{\rm K}$, $10^{-7}$ and $5\times 10^{-5}$, respectively
(with the self-shielding factor $f_{\rm sh}$ computed from
equations~\ref{eqn:fsh} and \ref{eqn:nh2}). These choices were motivated
by Figures \ref{fig:at4}-\ref{fig:bt5}, which show that $f_{\rm H_2}$
never exceeds $\sim 10^{-7}$ for the ``hot'' halos, and $f_{\rm e}$ is
between $10^{-5}$ and $10^{-4}$ at $n\sim 10^3$.  We therefore adopt
$10^{-7}$ as the critical ${\rm H_2}$ fraction, above which the gas
could cool below $\sim 8000~{\rm K}$.

The solid curves in Figures \ref{fig:explainT4} and
\ref{fig:explainT5} show the formation speeds $S_{\rm form}$, and the
other three curves show the photo-dissociation, collisional
dissociation, and total dissociation speeds, as labeled.  Note that at
low density, $S_{\rm form}$ increases with density, but at high
density, it stays approximately constant. This behavior follows from
equation~(\ref{eqn:kform}): in the low density limit, $k_{10}n\ll
k_{25}$, $k_{\rm form}\approx k_{9}k_{10}n/k_{25}\propto n$; whereas
in the high density limit, $k_{10}n\gg k_{25}$, $k_{\rm form}\approx
k_{9}\sim {\rm constant}$.  Likewise, the normalized dissociation rate
decreases with density when it is dominated by $S_{\rm pd}$, but
asymptotes to a constant in the high--density limit, where collisional
dissociation dominates.  Figures \ref{fig:explainT4} and
\ref{fig:explainT5} show that $S_{\rm cd}$ transits from a low value
to a high value around the critical density $n_{\rm cr}\sim 10^3~{\rm
cm^{-3}}$.

Most importantly, the formation and dissociation speeds are both
functions of $J$. In particular, as $J$ is increased, the total
dissociation rate is increased, but only below the density where ${\rm
H_2}$ photo--dissociation dominates collisional dissociation -- the
collisional rates are independent of $J$.  Likewise, as $J$ is
increased, the effective formation rate decreases due to $H^-$
photo--dissociation, but only at low densities, where ${\rm H^-}$
photo--dissociation is more important than collisional ${\rm H^-}$
dissociation (since, again, the collisional ${\rm H^-}$ dissociation
rate is independent of $J$).

The critical value, $J^{\rm crit}$, can be identified as the lowest
value for which the formation speed and the total dissociation speed
can become equal, for any value of the density.  If $J$ is lower than
this $J^{\rm crit}$, then, by definition, dissociation is always more
rapid than formation, and the ${\rm H_2}$ abundance can not increase
above the critical value.  However, if $J$ exceeds this $J^{\rm
crit}$, then at the density where $S_{\rm totd}=S_{\rm form}$, the
${\rm H_2}$ abundance increases, and the gas starts to cool. Because
$S_{\rm cd}$ drops further as the gas cools (see Fig. \ref{fig:k15}
below), more ${\rm H_2}$ molecules are able to form, and the cooling
proceeds in a runaway fashion.

Since the gas initially starts at low density, and, as shown in
Figures \ref{fig:explainT4} and \ref{fig:explainT5}, $S_{\rm totd}$ is
high at both low and high densities, whether or not a halo would cool
is mainly determined by the condition around the density $n_{\rm min}$
where $S_{\rm totd}$ reaches a minimum. The value at this minimum,
$n_{\rm min} \sim 10^2 -10^4~{\rm cm^{-3}}$ when $J_{21}$ is close to
$\jcrit$, coincides with the critical density for ${\rm H_2}$, $n_{\rm
cr}\approx 10^{3}~{\rm cm^{-3}}$ (see below).

Figures \ref{fig:explainT4} and \ref{fig:explainT5} also show the
relative importance of ${\rm H^-}$ photo-dissociation versus direct
${\rm H_2}$ photo-dissociation in setting the critical value of
$J_{21}$.  In particular, the $J$--dependence of $S_{\rm form}$ (solid
curves) arises from ${\rm H^-}$ photo-dissociation, whereas the
$J$--dependence of $S_{\rm totd}$ (dash-dotted curves) is from ${\rm
H_2}$ photo-dissociation.  Comparing the thick and thin solid curves
in Figures \ref{fig:explainT4} and \ref{fig:explainT5}, we see that
around the critical density of $\sim 10^3~{\rm cm^{-3}}$, ${\rm H^-}$
photo-dissociation is much more important in the T4 case. This is
primarily because of the softer shape of this spectrum: for a fixed
flux at $13.6$eV, the flux at $\sim 1$eV, just above the ${\rm H^-}$
photo--dissociation threshold, is much larger for the T4 than for the
T5 case (see Omukai et al. 2008 for an explicit comparison).
Interestingly, in the case of the T4 spectrum, as $J$ is varied, the
minimum in $S_{\rm totd}$ moves almost in parallel with the $S_{\rm
form}$ vs. $n$ curve. As a result, $\jcrit$ is, in fact, quite
insensitive to the direct ${\rm H_2}$ photo-dissociation rate (and
therefore also to the details of our treatment of self-shielding).  We
verified this conclusion explicitly, by artificially setting
$k_{28}=0$ in the one-zone model. In this case, $\jcrit$ only
increases by a factor of $\sim 2$.  The situation is very different in
the case of type T5 spectrum: near the critical density, $S_{\rm
form}$ is insensitive to $J$, and the critical value of $J_{21}$ is
determined almost entirely by ${\rm H_2}$ photo-dissociation.

As mentioned above, our $\jcrit$ values are a factor of 3-10 lower
than found by OM01 and other previous studies.  The one--zone models
can also be used to identify the reason for this discrepancy.  In
particular, we have systematically varied, one--by--one, each of the
five important reaction rates enumerated above, and re--computed
$\jcrit$ as these rates were varied.  We found that the difference in
$\jcrit$ is almost fully accounted for by the difference in our
adopted ${\rm H_2}$ collisional dissociation rate ($k_{15}$).  In
particular, when we change our $k_{15}$ to be the same as in the
one--zone calculations of OM01, but leave our other rates unchanged,
we find $J_{21}^{\rm crit}=1.1\times 10^3$ (for the type T4
background) and $3.4\times 10^4$ (for the type T5 background).  For
the T4 case, this new $\jcrit$ is very close to that found by OM01.
In the T5 case, our $\jcrit$ is still a factor of $\sim 3$ lower. We
have varied all of the other important reaction rates (i.e., setting
them to the values used in OM01), but found that this discrepancy
remains.  We suspect the remaining difference can be due to the
different cooling functions used.

In the Enzo runs, as well as in our one-zone models, we adopted the
${\rm H_2}$ collisional dissociation rate from Martin et al.  (1996),
while OM01 used a rate based on papers by Lepp \& Shull (1983),
Shapiro \& Kang (1987) and Palla et al. (1983). In Figure
\ref{fig:k15}, we show $k_{15}$ used in OM01 and Enzo, as a function
of temperature, for ${\rm H}$ densities of $10^{-3}~{\rm cm^{-3}}$
(low density), $10^3~{\rm cm^{-3}}$ and $10^9~{\rm cm^{-3}}$(high
density). As this figure shows, the rate used in Enzo is higher than
the rate used in OM01, especially at the intermediate density, which
is close to the critical density, and where the difference is more
than a factor of $10$. This large difference at the intermediate
density is primarily due to the difference in $n_{\rm cr}$. OM01
adopted $n_{\rm cr}$ from Lepp \& Shull (1983), which considered only
vibrational transitions.  Martin et al. (1996) takes into account all
rotational and vibrational-rotational transitions, and suggest that
accounting for both types of transitions reduces $n_{\rm cr}$ by a
factor of $\sim 10$.  This then explains and justifies the reason that
our $\jcrit$ values are lower than in OM01.

\begin{figure}
\begin{tabular}{c}
\rotatebox{-0}{\resizebox{90mm}{!}{\includegraphics{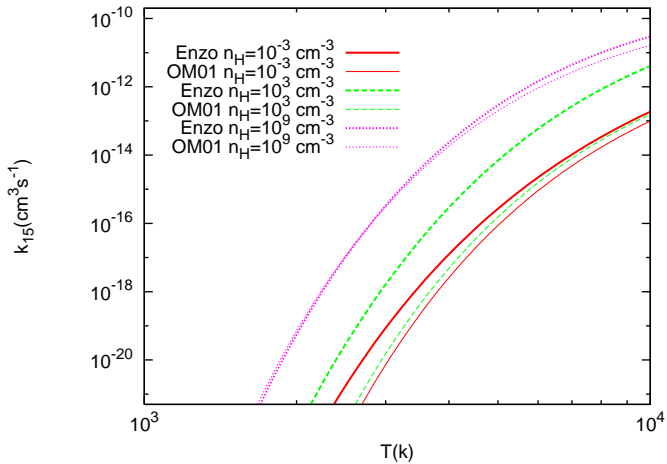}}}
\end{tabular}
\caption{The ${\rm H_2}$ collisional dissociation rate ($k_{15}$)
  adopted by Enzo and OM01, as a function of temperature, at three
  different densities. {\it Thick curves}: rates from Martin et
  al. (1986), used by Enzo; {\it Thin curves}: rates used by
  OM01. Solid, dashed and dotted curves show the rates at $n_{\rm
    H}=10^{-3}, 10^{3}$ and $10^{9}~{\rm cm^{-3}}$, respectively.}
\label{fig:k15}
\end{figure}
\subsection{The Origin of Scatter in $\jcrit$}
\label{subsec:scatter}

Another interesting result from the Enzo runs is that $J_{21}^{\rm
crit}$ found in halos B and C are higher than both in halo A and in
the one-zone model. This might be due to variations in the gas
temperatures in the collapsing halos. During the assembly of the
halos, due to the lack of spherical symmetry and variations in the
merger histories, shocks occur at various densities and Mach
numbers. This is known to cause a $\sim 20$\% scatter in the
temperature at fixed density (e.g. Loken et al. 2002).  The comparison
of Figures \ref{fig:at4} and \ref{fig:at5} with Figures \ref{fig:bt4}
and \ref{fig:bt5} shows that the temperature at $n_{\rm cr}$ in halo A
is always higher than in halo C for the $J_{21}$'s at which these two
halos differ ($J_{21}=10^2$ for type T4 background, $J_{21}=3\times
10^4$ for type T5 background). Among the five most important reactions
discussed above, ${\rm H_{2}}$ collisional dissociation is most
sensitive to temperature.  If the temperature is lower, the
collisional dissociation rate is also lower, which requires a higher
$\jcrit$ to compensate.

We used the one--zone model (which, recall, gives good agreement with
halo A) to check that this is the main reason for the different
$J^{\rm crit}$ values for halos A and C. We find that $J_{21}^{\rm
crit}$ indeed becomes slightly larger than $10^2$ ($3.2\times 10^4$)
in the case of a T4 (T5) background if the temperature is artificially
decreased by $\approx 20\%$, from $\approx 8,000~{\rm K}$ to
$6,000~{\rm K}$ at a density of $5\times 10^2~{\rm cm^{-3}}$ ($5\times
10^3~{\rm cm^{-3}}$). Note that the temperature difference occurs at a
higher density in the case of a T5 background.  $\jcrit$ could not
reach $3\times 10^4$ if we set the temperature to $6000~{K}$ at
density lower than $\sim 3\times 10^3~{\rm cm^{-3}}$. This is
consistent with Figures \ref{fig:bt4} and \ref{fig:bt5}, where we see
that the temperature drops at larger radius (lower density) in the
case of $J_{21}=10^2$ and the T4 spectrum than in the case of
$J_{21}=3\times 10^4$ and the T5 spectrum.  Figures
\ref{fig:explainT4} and \ref{fig:explainT5} also show that the minimum
in the total dissociation rate occurs at a higher density in the case
of the T5 spectrum. As Figure \ref{fig:k15} shows, $k_{15}$ decreases
significantly -- by about one order of magnitude -- when the
temperature is lowered to from $8000~{K}$ to $6000~{K}$.

The ${\rm H_2}$ formation rate ($k_{10}$) and the ${\rm H^-}$
formation rate ($k_9$) also dependent on temperature, but they work in
the wrong direction to explain the increased $\jcrit$ for halo C:
$k_9$ and $k_{10}$ are lower when the temperature is lower, which
would tend to reduce $\jcrit$.  However, the temperature dependence of
$k_9$ and $k_{10}$ are both much weaker than that of $k_{15}$. When
the temperature is changed from 8000 K to 6000 K, $k_9$ and $k_{10}$
decrease by 23\% and 5\%, respectively.  We conclude that random
temperature variations at the level of $\sim 20\%$, which naturally
occur due to variations in the strengths of shocks that occur in the
collapsing gas, can account for the scatter we observed in $J^{\rm
crit}$.

\subsection{${\rm H_2}$ Self-Shielding}
\label{subsec:selfshielding}

The value of $\jcrit$ is directly related to the self-shielding factor
$f_{\rm sh}$ that we computed with an approximate method (equations
\ref{eqn:fsh} and \ref{eqn:nh2}). We here check the accuracy of these
approximations a--posteriori, using the simulation outputs. We
computed $N_{\rm H_2}$ and $f_{\rm sh}$ by integrating the ${\rm H_2}$
profile from the outside in, and compared these values with those
obtained from Equations \ref{eqn:fsh} and \ref{eqn:nh2}.

\begin{figure}
\begin{tabular}{c}
\rotatebox{-0}{\resizebox{90mm}{!}{\includegraphics{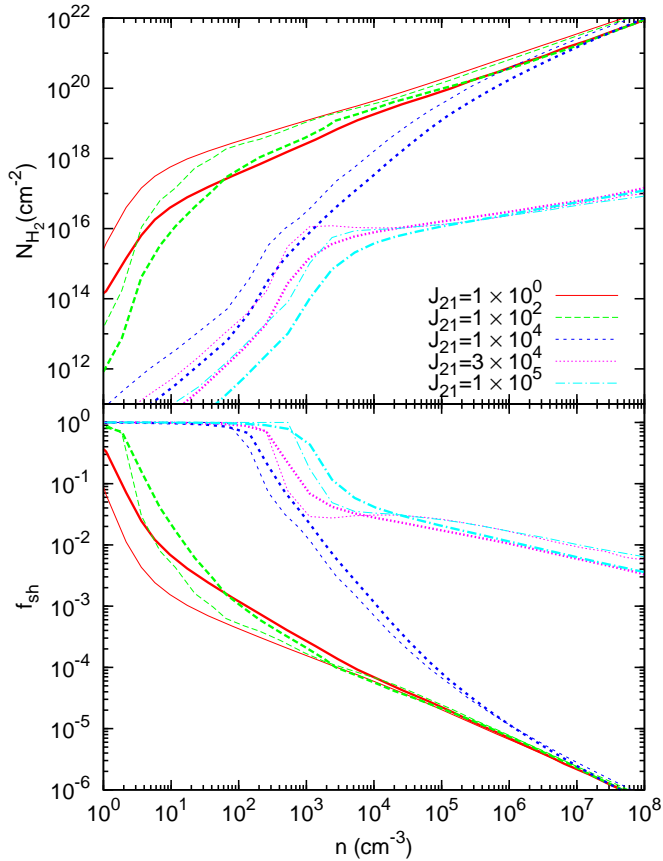}}}
\end{tabular}
\caption{The ${\rm H_2}$ column density and self-shielding factor in
  the Lyman--Werner bands for Halo A and a T5 background. The {\it
  thin curves} were computed from the approximations in Equations
  \ref{eqn:fsh} and \ref{eqn:nh2}; the {\it thick curves} were
  obtained by numerically integrating the ${\rm H_2}$ profile in the
  simulations.  The various line types correspond to different
  intensities $J$, as labeled.  See text for details.}
\label{fig:nh2fsh}
\end{figure}

We found results that were qualitatively similar in all the cases, and
show only one example of such a comparison in Figure \ref{fig:nh2fsh};
for Halo A in the presence of a T5 background. The thick (thin) curves
show the results from the numerical integration (approximate
method). Since the halos are not spherically symmetric, integrating
along different sight--lines gives different results. In our numerical
integration, we first obtain a spherically average halo profile from
the simulations outputs. At a given density $n$, we then obtain
$f_{\rm sh}$, by averaging the self-shielding factor, over all
directions, at the radius with this density.  For $N_{\rm H_2}$,
simple averaging makes little sense (the photodissociation rate is
linearly proportional to $f_{\rm sh}$, but not to $N_{\rm H_2}$), so
we only show the results of integrating along the single radial
sight--line in the direction away from the halo center.

Figure \ref{fig:nh2fsh} shows that over the interesting density range,
$N_{\rm H_2}$ and $f_{\rm sh}$ obtained from Equations \ref{eqn:fsh}
and \ref{eqn:nh2} agree within a factor $\sim 10$, with the values
obtained from the non--local integrations. In those cases when the
halo gas cools, $N_{\rm H_2}$ computed with the approximate method is
always larger, by a factor of a few, while in those cases when the
halo stays hot, $N_{\rm H_2}$ computed with the approximate method is
larger at low density, but becomes very close to the non--local value
at densities above $\ge 10^5~{\rm cm^{-3}}$. This is explained simply
by the fact that Equation \ref{eqn:nh2} assumes that the ${\rm H_2}$
fraction is constant.  As Figures \ref{fig:at4} -- \ref{fig:bt5} show,
this is a poor assumption for ``hot'' halos and in the outer region of
``cool'' halos, where $f_{\rm H_2}$ decreases with radius. Only in the
inner region ($R\lsim 10~{\rm pc}$) of ``cool'' halos does $f_{\rm
H_2}$ become approximately constant ($\sim 10^{-8}$).  From any given
location, the sightline directly pointing outward (away from the halo
center) is typically the least self-shielded, and the mean non--local
self-shielding factor is therefore lower than one would obtain
assuming a spatially constant $f_{\rm H_2}$ (except close to the halo
center).  As a result, the thick curves in the $f_{\rm sh}$ panel of
Figure \ref{fig:nh2fsh} are offset downward from the positions
expected from the $N_{\rm H_2}$ panel using Equation \ref{eqn:fsh}.

Since at ${\rm n_{cr}}$, $f_{\rm H_2}$ computed with the approximate
method is always slightly lower than its non--local value, we expect
that $\jcrit$ would be slightly lower if $f_{\rm H_2}$ were
computed exactly.  Furthermore, self-shielding could be overestimated
in our treatment, since we have ignored bulk motions of the gas.  In
the extreme limit where we set the ${\rm H_2}$ self-shielding factors
to zero, in the one--zone model, we find that the critical fluxes are
reduced by factors of $\approx 2$ and $\approx 30$, to $J_{21}^{\rm
  crit}=19$ and $\jcrit=440$, for the type T4 and T5
backgrounds, respectively.

\section{The Mass of the Central Object}
\label{sec:fate}

The thermodynamical properties of the collapsing gas have a large
impact on the final object that forms in the core of the halo.  An
estimate for the mass of the central object, in particular, can be
obtained by the following simple argument, under the assumption that
the object is a single (super-)massive star. There exists a radius at
which the mass accretion time--scale $t_{\rm acc}$ equals the
Kelvin-Helmholtz time scale $t_{\rm KH}$ for a proto--star, with the
proto--stellar mass equal to the gas mass enclosed within this radius.
The gas outside this radius does not have the time to be incorporated
onto the proto--star before it settles to the zero--age main sequence
(ZAMS); conversely, the mass inside this radius will accrete onto the
protostar before the star settles to the ZAMS (e.g. Abel et al. 2002;
Omukai \& Palla 2003; O'Shea \& Norman 2007).  While this argument
ignores various feedback processes that can occur and affect the final
stellar mass, it gives a useful order--of--magnitude estimate.  See
McKee \& Tan (2008) for a more detailed discussion in the context of
first star formation in minihalos.

\begin{figure}
\begin{tabular}{c}
\rotatebox{-0}{\resizebox{90mm}{!}{\includegraphics{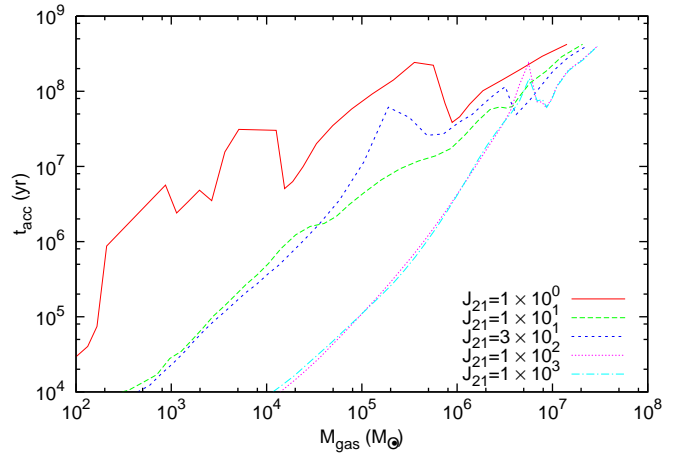}}}
\end{tabular}
\caption{The local accretion time--scale $t_{\rm acc}$ as a function
  of the enclosed gas mass $M_{\rm gas}$ for Halo A and a T4
  background with different intensities, as labeled.}
\label{fig:acc1}
\end{figure}

\begin{figure}
\begin{tabular}{c}
\rotatebox{-0}{\resizebox{90mm}{!}{\includegraphics{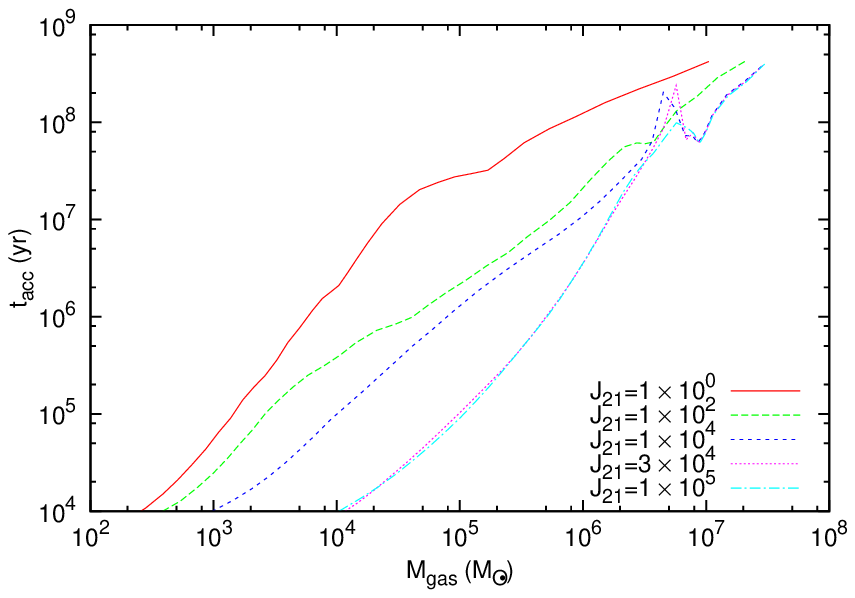}}}
\end{tabular}
\caption{Same as Figure \ref{fig:acc1} except for type T5
backgrounds.}
\label{fig:acc2}
\end{figure}

In Figures \ref{fig:acc1} and \ref{fig:acc2}, we show the accretion
time scale $t_{\rm acc}\equiv R/\langle v_{R}\rangle$ as a function of
the gas mass $M_{\rm gas}$ enclosed within a sphere of radius
$R$. Here $\langle v_{R}\rangle$ is the mean radial in-fall velocity
at $R$. These figures show that at enclosed masses of $<10^6~{\rm
M_\odot}$, the accretion rates in the ``hot'' halos are 1-2 orders of
magnitude higher than for the ``cool'' halos.  The Kelvin-Helmholtz
contraction time scale for metal--free stars is approximately $10^5$
years, with a relatively weak dependence on the proto--stellar mass
(Schaerer 2002).  Within this time, the central object in a ``hot''
halo will accumulate $10^5~\msun$, while the object in a ``cool'' halo
could accumulates $10^2-10^4~\msun$.  The latter values are very
similar to previous estimates for the masses of the first stars that
form in minihalos (Abel et al. 2002; Bromm et al. 2002; O'Shea \&
Norman 2007); this is not surprising, given that the densities and
temperatures reached in these ``cool'' halos via efficient ${\rm
H_2}$--cooling are very similar to those in lower--mass minihalos.
However, the value $10^5~\msun$ for the ``hot'' halos is much larger;
in particular, a supermassive, metal--free star with this large a mass
suffers from post--Newtonian instabilities, and ultimately collapses
without an explosion to produce a SMBH of the same mass (Fuller,
Woosley \& Weaver 1986).

\begin{figure}
\begin{tabular}{c}
\rotatebox{-0}{\resizebox{90mm}{!}{\includegraphics{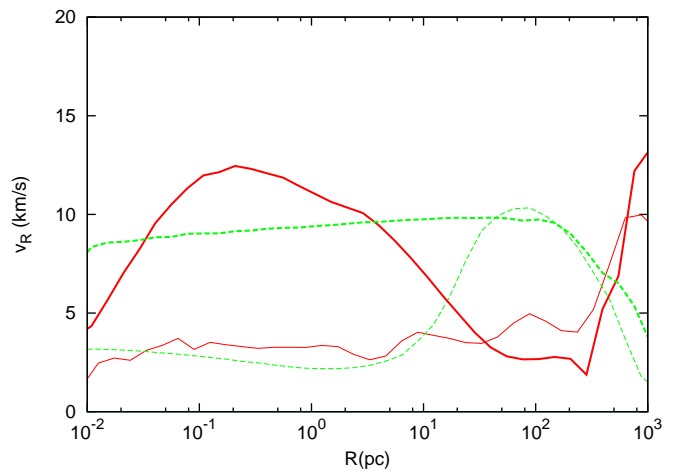}}}
\end{tabular}
\caption{The radial velocity ({\it solid curves}) and sound speed
    ({\it dashed curves}) as a function of radius in halo A in the
    presence of type T4 backgrounds. The thick and thin curves are for
    $J_{21}=10^1$ and $10^3$, respectively.}
\label{fig:vr}
\end{figure}

\begin{figure}
\begin{tabular}{c}
\rotatebox{-0}{\resizebox{90mm}{!}{\includegraphics{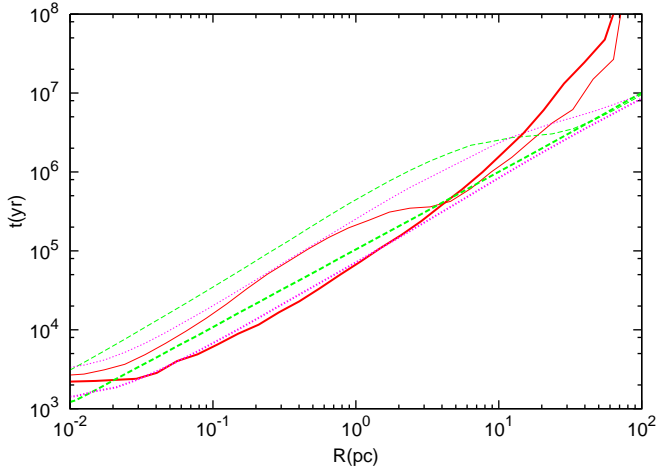}}}
\end{tabular}
\caption{The cooling ({\it solid curves}), sound--crossing ({\it
  dashed curves}) and free--fall time scale ({\it dotted curves}) as a
  function of radius in halo A, in the presence of type T4
  backgrounds. The thick and thin curves are for $J_{21}=10^1$ and
  $10^3$, respectively.}
\label{fig:tscale}
\end{figure}

Given that the effect of the UV background is to dissociate molecules,
and to reduce the efficiency of cooling, it may be surprising that the
gas accretion rate in the ``hot'' halos is higher than in the ``cool''
halos, whose gas cools more efficiently.  However, we note that the
accretion rates in Figure~\ref{fig:acc1} and \ref{fig:acc2} are shown
at different times: although the gas collapse in the ``hot'' halos is
more rapid, it occurs after a significant delay.  Once the collapse
begins, it proceeds over the dynamical time scale,
\begin{eqnarray}
\label{eqn:taccr0}
t_{\rm acc}\sim t_{\rm ff}=\sqrt{\frac{3\pi}{32G\rho}}.
\end{eqnarray}
Here, $G$ is the gravitational constant, and $\rho$ is the total (gas
+ dark matter) density.  On the other hand, the infall speed is
modulated by the sound speed. Once the infall becomes supersonic, weak
shocks occur, which tend to slow the infall.  These shocks also
provide an additional source of heating the gas, thereby elevating the
sound speed.  As a result, the sound speed and the infall speed tend
to trace each other.  In Figure \ref{fig:vr}, we show spherically
averaged profiles of the radial velocity (solid curves) and the sound
speed (dashed curves) in halo A (and type T4 backgrounds). The thick
and thin curves are for $J_{21}=10^1$ and $J_{21}=10^3$, respectively.
Although not exactly equal, the sound speed and the radial infall
speed are always of the same.
A similar conclusion could be drawn from Figure \ref{fig:tscale},
where we show three time scales: the cooling time (solid curves), the
sound crossing time (defined as $R/c_s$; dashed curves) and the
free--fall time (dotted curves).  The thick and thin curves are for
the same two values of $J_{21}$ as in Figure \ref{fig:vr}. Inside the
virial radius (i.e., within a few pc, where the gas is first shocked
and heated to $10^4$K) the cooling and sound crossing times both trace
the free--fall time scale to within a factor of $\sim 2$.  It is
particularly revealing that the good match between the three
time--scales holds over a factor of $\sim 300$ in radii.  Using the
condition that the sound crossing time equals the free--fall time,
\begin{eqnarray}
\label{eqn:vtheorem}
R/c_{\rm sound}=R\sqrt{\frac{\mu m_{\rm H}}{\gamma kT}}\approx \sqrt{\frac{3\pi}{32G\rho}},
\end{eqnarray}
together with $M\sim \rho R^3$ to convert $\rho$ in Equation
\ref{eqn:taccr0} to $M$ and $T$, we find
\begin{eqnarray}
\label{eqn:taccr}
t_{\rm accr}\approx t_{\rm ff}\approx GM\left(\frac{\mu m_{\rm H}}{\gamma
  kT}\right)^{3/2}=\frac{GM}{c_{\rm sound}^{3}},
\end{eqnarray}
where a numerical factor of order unity (which depends on the actual
density profile) has been omitted.  Note that to within these factors,
this result agrees with the well--known result for the accretion
time--scale in the self--similar collapse of a singular isothermal
sphere (Shu 1977).
In their simulations, O'Shea \& Norman (2007) find the same scaling of
the accretion rate with sound--speed, and attribute this scaling to
the physics of the collapsing isothermal sphere; note that our
explanation -- namely that weak shocks limit the infall velocity to be
close to the sound speed -- is somewhat different.  Equation
\ref{eqn:taccr} predicts a linear dependence between $t_{\rm accr}$
and the enclosed mass $M$, as well as $t_{\rm acc}\propto T^{-3/2}$;
both of these scalings are in good agreement with the accretion times
shown in Figures \ref{fig:acc1} and \ref{fig:acc2}. As discussed in
\S~\ref{sec:results} above, the central gas temperatures in the
``hot'' halos are a factor of $\sim 6$ higher than those in the
``cool'' halos (with the exception of the runs with $J_{21}\le 1$).
Based on the above argument, $t_{\rm acc}$ is therefore expected to be
a factor of $\sim 15$ smaller, at a given $M_{\rm gas}$, for the
``hot'' halos. This factor is indeed in good agreement with the
difference in the accretion rates for the two types of halos shown in
Figures \ref{fig:acc1} and \ref{fig:acc2}.

\section{Conclusions}
\label{sec:conclusion}

By performing a series of simulations with the AMR code Enzo, we
determined the critical intensity of the UV background needed to
suppress the ${\rm H_2}$ cooling in dark matter halos with virial
temperatures of $T_{\rm vir}\gsim 10^4$K.  We have modified Enzo to
include ${\rm H^-}$ photo-dissociation, as well as self--shielding of
${\rm H_2}$ in the Lyman--Werner bands. To increase the computing
speed, the self--shielding effect was calculated with an approximate
method that uses only local physical quantities; we have shown,
however, that this simple approximation is accurate to better than an
order of magnitude.

In agreement with earlier results from a one--zone model (OM01), we
found that, depending on whether the background intensity is above or
below the critical value, the halos can be unambiguously grouped into
two categories: ``hot'' halos and ``cool'' halos. In general, ``hot''
halos have much higher central gas temperature, density, and electron
fraction and a much lower ${\rm H_2}$ fraction. In the ``cool'' halos,
the temperature drops inside $R\sim 10$ pc, which corresponds to a
local density $n\sim 10^3~{\rm cm^{-3}}$.

For type T4 (T5) spectra, $\jcrit$ is found to be in the range of
$3\times 10^1$ -- $10^2$ ($10^4$ -- $3\times 10^4$) for halo A, and
$10^2$ -- $2\times 10^2$ ($3\times 10^4$ -- $10^5$) for halos B and
C. These values show that the critical flux varies significantly from
halo to halo; we attribute this variation to scatter in temperature at
a given density.  Most importantly, the values of $\jcrit$ we find are
a factor of 3--10 lower than previously estimated.  Using one-zone
models with simplified dynamics, we studied the dependence of $\jcrit$
on each of the important reaction rates. We have shown that the
difference between our results and previous estimates can be
attributed to our adoption of a different, more accurate, ${\rm H_2}$
collisional dissociation rate.

In those halos in which ${\rm H_2}$ cooling is suppressed, the gas
cools efficiently, but remains relatively hot, at a temperature near
$T\sim 8000$ K.  While gas collapse starts with a significant delay,
we have shown that, as a result of the elevated temperature, the gas
accretion rate in these halos is ultimately increased by $\approx 1-2$
orders of magnitude. As a result, a supermassive star with a mass of
$10^5~{\rm M_\odot}$ may form in the cores of these halos (compared to
$10^{2-3}~{\rm M_\odot}$ stars in the presence of ${\rm H_2}$
cooling), ultimately producing a supermassive black hole (SMBH) with a
comparable mass.

The critical UV fluxes required to suppress ${\rm H_2}$ cooling in
halos with $T_{\rm vir}\gsim 10^4$K are high compared to the expected
level of the cosmic UV background at high redshifts. Most likely, the
halos exposed to a super--critical UV flux are a small subset of all
$T_{\rm vir}\gsim 10^4$K halos that happen to sample the bright--end
tail of the fluctuating cosmic UV background.  This makes the
reductions in $\jcrit$ that we have found particularly significant:
according to a model for the fluctuating UV background, as sampled by
DM halos (Dijkstra et al. 2008) the 10--fold decrease in $J^{\rm
crit}$ can increase the number of candidate DM halos, where direct
SMBH formation may be feasible, by a factor of $\approx 10^3$.

\vspace{-0.5\baselineskip}

\section*{Acknowledgments}

This work was supported by the NSF grant AST-05-07161.  ZH
acknowledges support by the Pol\'anyi Program of the Hungarian
National Office of Technology.  GB acknowledges support from NSF
grants AST-05-07161, AST-05-47823, and AST-06-06959, as well as
computational resources from the National Center for Supercomputing
Applications.


\onecolumn
\centerline{APPENDIX: REACTION RATES AND CROSS SECTIONS}

\begin{center}
\begin{table}
\label{tbl:reactions}
\begin{tabular}{l l l}
\hline\hline
Reaction & Rate Coefficient $k$ (${\rm cm^3~s^{-1}}$) or Cross-section $\sigma$ (${\rm cm^2}$) &\\
\hline
(1) ${\rm H +  e^-    \rightarrow H^+ +2e^-  }$ & 
$ k_1={\rm exp}(-32.71396786375 $
& \\
&$      + 13.53655609057 ({\rm ln}T_{\rm eV}) $&\\
&$      - 5.739328757388 ({\rm ln}T_{\rm eV})^2 $&\\
&$      + 1.563154982022 ({\rm ln}T_{\rm eV})^3 $&\\
&$      - 0.2877056004391 ({\rm ln}T_{\rm eV})^4 $&\\
&$      + 0.03482559773736999 ({\rm ln}T_{\rm eV})^5 $&\\
&$      - 0.00263197617559 ({\rm ln}T_{\rm eV})^6 $&\\
&$      + 0.0001119543953861 ({\rm ln}T_{\rm eV})^7 $&\\
&$      - 2.039149852002\times 10^{-6} ({\rm ln}T_{\rm eV})^8) $&\\

(2) ${\rm He      +  e^-    \rightarrow He^+    +2e^-  }$ &
$k_2 = {\rm exp}(-44.09864886561001$
& \\
&$      + 23.91596563469 ({\rm ln}T_{\rm eV}) $&\\
&$      - 10.75323019821 ({\rm ln}T_{\rm eV})^2 $&\\
&$      + 3.058038757198 ({\rm ln}T_{\rm eV})^3 $&\\
&$      - 0.5685118909884001 ({\rm ln}T_{\rm eV})^4 $&\\
&$      + 0.06795391233790001 ({\rm ln}T_{\rm eV})^5 $&\\
&$      - 0.005009056101857001 ({\rm ln}T_{\rm eV})^6 $&\\
&$      + 0.0002067236157507 ({\rm ln}T_{\rm eV})^7 $&\\
&$      - 3.649161410833\times 10^{-6} ({\rm ln}T_{\rm eV})^8) $&\\

(3) ${\rm He^+    +  e^-    \rightarrow He^{++}  +2e^-  }$ &
$         k_3 = {\rm exp}(-68.71040990212001$
&  \\ 
&$      + 43.93347632635 ({\rm ln}T_{\rm eV}) $&\\
&$      - 18.48066993568 ({\rm ln}T_{\rm eV})^2 $&\\
&$      + 4.701626486759002 ({\rm ln}T_{\rm eV})^3 $&\\
&$      - 0.7692466334492 ({\rm ln}T_{\rm eV})^4 $&\\
&$      + 0.08113042097303 ({\rm ln}T_{\rm eV})^5 $&\\
&$      - 0.005324020628287001 ({\rm ln}T_{\rm eV})^6 $&\\
&$      + 0.0001975705312221 ({\rm ln}T_{\rm eV})^7 $&\\
&$      - 3.165581065665\times 10^{-6} ({\rm ln}T_{\rm eV})^8) $&\\

(4) ${\rm H^+     +  e^-    \rightarrow H       + h\nu }$ &
$         k_4= {\rm exp}(-28.61303380689232$
&   \\
&$   - 0.7241125657826851 ({\rm ln}T_{\rm eV}) $&\\
&$   - 0.02026044731984691 ({\rm ln}T_{\rm eV})^2 $&\\
&$   - 0.002380861877349834 ({\rm ln}T_{\rm eV})^3 $&\\
&$   - 0.0003212605213188796 ({\rm ln}T_{\rm eV})^4 $&\\
&$   - 0.00001421502914054107 ({\rm ln}T_{\rm eV})^5 $&\\
&$   + 4.989108920299513\times 10^{-6} ({\rm ln}T_{\rm eV})^6 $&\\
&$   + 5.755614137575758\times 10^{-7} ({\rm ln}T_{\rm eV})^7 $&\\
&$   - 1.856767039775261\times 10^{-8} ({\rm ln}T_{\rm eV})^8 $&\\
&$   - 3.071135243196595\times 10^{-9} ({\rm ln}T_{\rm eV})^9)  $&\\

(5) ${\rm He^+    +  e^-    \rightarrow He      + h\nu }$ &
$         k_5 = 1.54\times 10^{-9} (1+0.3/{\rm exp}(8.099328789667/T_{\rm eV}))$
&  \\ 
&$      / ({\rm exp}(40.49664394833662/T_{\rm eV})\times T_{\rm eV}^{1.5}) $&\\
&$      + 3.92\times 10^{-13}/T_{\rm eV}^{0.6353}  $&\\

(6) ${\rm He^{++} +  e^-    \rightarrow He^+    + h\nu }$ &
$k_6 = 3.36\times 10^{-10} T^{-\frac{1}{2}} (T/1000)^{-0.2} (1+(T\times 10^{-6})^{0.7})^{-1}
$                                 
&  \\ 

(7) ${\rm H       +  H^+    \rightarrow H_2^+   + h\nu }$ &  
$k_7 = 1.85\times 10^{-23}\times T^{1.8} $
& $T\le 6.7\times 10^3~{\rm K}$ \\ 
&$   k_7 = 5.81\times 10^{-16} (T/56200)^{({-0.6657} {\rm
  log}(T/56200))} $& $T > 6.7\times 10^3~{\rm K}$\\

\hline
\end{tabular}
\end{table}
\end{center}

\begin{center}
\begin{table}
\label{tbl:reactions2}
\begin{tabular}{l l l}
\hline\hline

Reaction & Rate Coefficient $k$ (${\rm cm^3~s^{-1}}$) or Cross-section $\sigma$ (${\rm cm^2}$) &\\
\hline

(8) ${\rm H_2^+   +  H      \rightarrow H_2     + H^+  }$ &
$6.0\times 10^{-10}                               $      &  \\ 

(9) ${\rm H       +  e^-    \rightarrow H^-  + h\nu }$ & 
$k_{9} = 6.775\times 10^{-15} T_{\rm eV}^{0.8779}$
&  \\ 

(10) ${\rm H       +  H^-    \rightarrow H_2     + e^-  }$ &
$k_{10}=1.43\times 10^{-9}                                $
&$T_{\rm eV}\le 0.1$ \\
&$k_{10}= {\rm exp}(-20.06913897587003$& $T_{\rm eV}>0.1$\\
&$   + 0.2289800603272916 ({\rm ln}T_{\rm eV}) $&\\
&$   + 0.03599837721023835 ({\rm ln}T_{\rm eV})^2 $&\\
&$   - 0.004555120027032095 ({\rm ln}T_{\rm eV})^3 $&\\
&$   - 0.0003105115447124016 ({\rm ln}T_{\rm eV})^4 $&\\
&$   + 0.0001073294010367247 ({\rm ln}T_{\rm eV})^5 $&\\
&$   - 8.36671960467864\times 10^{-6} ({\rm ln}T_{\rm eV})^6 $&\\
&$   + 2.238306228891639\times 10^{-7} ({\rm ln}T_{\rm eV})^7) $&\\

(11) ${\rm H_2^+   +  e^-    \rightarrow 2H              }$ &
$k_{11}=1.0\times 10^{-8}        $   & $T\le 617~{\rm K}$\\
&$   k_{11} = 1.32\times 10^{-6} T^{-0.76} $&$T> 617~{\rm K}$\\
 
(12) ${\rm H_2^+   +  H^-    \rightarrow H_2     + H    }$ &
$k_{12}=5.00\times 10^{-6} T^{-\frac{1}{2}}               $   &  \\ 

(13) ${\rm H^-     +  H^+    \rightarrow 2H            }$ &
$k_{13} =6.5\times 10^{-9} T_{\rm eV}^{-\frac{1}{2}}$
&  \\ 

(14) ${\rm H_2     +  e^-   \rightarrow H        + H^-  }$ &
$k_{14}=0.0$ (not used in Enzo)
&  \\ 

(15) ${\rm H_2     +  H     \rightarrow 3H     } $ &
see expression in Martin et al. (1996)
&  \\ 

(16) ${\rm H_2     +  H_2   \rightarrow H_2     +2H     }$ & 
$k_{16}=0.0  $ (not used in Enzo)                & \\ 

(17) ${\rm H_2     +  H^+   \rightarrow H_2^+  + H     }$ &  
$k_{17}= {\rm exp}(-24.24914687731536$
&  \\
&$   + 3.400824447095291 ({\rm ln}T_{\rm eV}) $&\\
&$   - 3.898003964650152 ({\rm ln}T_{\rm eV})^2 $&\\
&$   + 2.045587822403071 ({\rm ln}T_{\rm eV})^3 $&\\
&$   - 0.5416182856220388 ({\rm ln}T_{\rm eV})^4 $&\\
&$   + 0.0841077503763412 ({\rm ln}T_{\rm eV})^5 $&\\
&$   - 0.007879026154483455 ({\rm ln}T_{\rm eV})^6 $&\\
&$   + 0.0004138398421504563 ({\rm ln}T_{\rm eV})^7 $&\\
&$   - 9.36345888928611\times 10^{-6} ({\rm ln}T_{\rm eV})^8) $&\\
 
(18) ${\rm H_2     +  e^-   \rightarrow 2H      +e^-    }$ &
$k_{18}=5.6\times 10^{-11} {\rm exp}(-102124/T) T^{0.5} $
&  \\

(19) ${\rm H^-     +  e^-   \rightarrow H       +2e^-   }$ &
$         k_{19} = {\rm exp}(-18.01849334273$
&  \\
&$      + 2.360852208681 ({\rm ln}T_{\rm eV}) $&\\
&$      - 0.2827443061704 ({\rm ln}T_{\rm eV})^2 $&\\
&$      + 0.01623316639567 ({\rm ln}T_{\rm eV})^3 $&\\
&$      - 0.03365012031362999 ({\rm ln}T_{\rm eV})^4 $&\\
&$      + 0.01178329782711 ({\rm ln}T_{\rm eV})^5 $&\\
&$      - 0.001656194699504 ({\rm ln}T_{\rm eV})^6 $&\\
&$      + 0.0001068275202678 ({\rm ln}T_{\rm eV})^7 $&\\
&$      - 2.631285809207\times 10^{-6} ({\rm ln}T_{\rm eV})^8)  $&\\
 
(20) ${\rm H^-     +  H     \rightarrow 2H      +e^-    }$ &
$k_{20} = 2.56\times 10^{-9} T_{\rm eV}^{1.78186} $
&  $T_{\rm eV}\le 0.1$\\ 
&$k_{20} = {\rm exp}(-20.37260896533324 $& $T_{\rm eV}>0.1$\\
&$      + 1.139449335841631 ({\rm ln}T_{\rm eV}) $&\\
&$      - 0.1421013521554148 ({\rm ln}T_{\rm eV})^2 $&\\
&$      + 0.00846445538663 ({\rm ln}T_{\rm eV})^3 $&\\
&$      - 0.0014327641212992 ({\rm ln}T_{\rm eV})^4 $&\\
&$      + 0.0002012250284791 ({\rm ln}T_{\rm eV})^5 $&\\
&$      + 0.0000866396324309 ({\rm ln}T_{\rm eV})^6 $&\\
&$      - 0.00002585009680264 ({\rm ln}T_{\rm eV})^7 $&\\
&$      + 2.4555011970392\times 10^{-6} ({\rm ln}T_{\rm eV})^8 $&\\
&$      - 8.06838246118\times 10^{-8} ({\rm ln}T_{\rm eV})^9) $&\\

(21) ${\rm H^-     +  H^+   \rightarrow H_2^+   +e^-    }$ & 
$      k_{21}=4\times 10^{-4} T^{-1.4} {\rm exp}(-15100/T)$
&  $T\le 10^4~{\rm K}$\\ 
&$k_{21}=1\times 10^{-8} T^{-0.4} $& $T> 10^4~{\rm K}$\\
%

(22) ${\rm H       +h\nu    \rightarrow H^+  +e^-   }$ &
$\sigma_{22}=6.3\times10^{-18}(\frac{\rm
  13.6eV}{h\nu})^4\exp(4-4(\tan^{-1})\epsilon/\epsilon)/[1-\exp
  (-2\pi/\epsilon)] ; \epsilon\equiv\sqrt{\frac{h\nu}{\rm 13.6}-1} $
&  \\ 
(23) ${\rm He      +h\nu    \rightarrow He^+  +e^-   }$ &
$\sigma_{23}=0.694\times10^{-18} [(\frac{h\nu}{\rm eV})^{1.82} + (\frac{h\nu}{\rm
    eV})^{3.23}]^{-1}$                                  &  \\ 
(24) ${\rm He^+    +h\nu    \rightarrow He^{++} +e^-   }$ &
$\sigma_{24}=1.575\times10^{-18}(\frac{\rm
  54.4eV}{h\nu})^4\exp(4-4(\tan^{-1})\epsilon/\epsilon)/[1-\exp
  (-2\pi/\epsilon)] ; \epsilon\equiv\sqrt{\frac{h\nu}{54.4}-1} $
&  \\ 
(25) ${\rm H^-     +h\nu    \rightarrow H  +e^-   }$ &
$k_{25}=10^{-10}\alpha J_{21}$; $\alpha=2000$ for T4 spectrum, $0.1$ for T5 spectrum & \\ 
(26) ${\rm H_2^+   +h\nu    \rightarrow H  +H^+   }$ &     {\rm see
  expression in Shapiro \& Kang (1987)}    &  \\ 
(27) ${\rm H_2     +h\nu    \rightarrow H_2^+   +e^-   }$ &  {\rm see
  expression in Shapiro \& Kang (1987)}                      & \\ 
(28) ${\rm H_2     +h\nu    \rightarrow 2H     }$
&
$k_{28}=10^{-12}\beta J_{21}$; $\beta=3$ for T4 spectrum, $0.9$ for T5 spectrum
& \\ 

\hline
\end{tabular}

\end{table}
\end{center}
{\footnotesize $T$ and $T_{\rm eV}$ are the gas temperature in units
  of K and eV, respectively.}

\vspace{\baselineskip} 


\label{lastpage}
\end{document}